\documentclass[12pt]{iopart}
\usepackage{graphicx,color,epsfig,rotate}
\begin{document}
\title[
New insight into the physics of iron] 
{New insight into  
the physics
of iron pnictides 
from optical  
and penetration depth data}

\author{S-L Drechsler$^1$, H Rosner$^2$, M Grobosch$^1$,  
G~Behr$^1$, F~Roth$^1$, G~Fuchs$^1$, K Koepernik$^1$,
  R~Schuster$^1$, J~Malek$^{1,3}$, S~Elgazzar$^4$, M~Rotter$^5$,
  D~Johrendt$^5$, H-H~Klauss$^6$, B~B\"uchner$^1$, and M~Knupfer$^1$}
\address{$^1$ IFW Dresden, P.O. Box 270116, D-01171 Dresden , Germany}
\address{$^2$ Max-Planck-Institute for Chemical Physics of Solids, Dresden, 
Germany}
\address{$^3$ Institute of  Physics,  
ASCR, Prague, Czech Republic}
\address{$^{4}$ Dept.\ of Physics, Faculty of Science, 
Menoufia University,  
Shebin El-kom, Egypt} 

\address{$^5$ University of Munich, Dept.\ of Chem. \& Biochem.,
D-81377 Munich, Germany}
\address{$^6$ Institut f\"ur Festk\"orperphysik, TU Dresden, D-01069 
Dresden, 
Germany }
\ead{s.l.drechsler@ifw-dresden.de}

\begin{abstract}
We report theoretical values for the unscreened plasma frequencies
$\Omega_{p}$ of several Fe pnictides obtained
from density functional theory (DFT)
based
calculations within the local density approximation (LDA)
 and  compare them  with experimental plasma frequencies obtained from 
reflectivity
measurements on both polycrystalline samples
and single crystals. The sizable
renormalization observed for all considered compounds points to the 
presence
of significant many-body effects beyond the LDA. 
From the  large values of the empirical 
background dielectric constant $\varepsilon_{\infty} \approx $
12-15 derived from reflectivity data, 
we estimate a large arsenic polarizability 
$\tilde{\alpha}_{\rm As} \approx 9.5 \pm 1.2$~\AA$^3$ where the details 
depend on  the polarizabilities of the remaining ions taken 
from the literature.
This large polarizability can significantly
reduce the value of the Coulomb repulsion
$U_d\sim$4~eV on iron known from iron oxides to a level of 2~eV or below. 
In general, independently on such details, this result 
points to rather strong polaronic effects 
as suggested by G.A.~Sawatzky {\it et al.}, in 
references {\it arXiv}:0808.1390 
\cite{sawatzky} and {\it arXiv}:0811.0214\cite{berciu}. 
Possible consequences for the conditions of a formation of bipolarons
are discussed, too. 
From the extrapolated
$\mu$SR (muon spin rotation) penetration
depth data at very low-temperature and the
experimental value of the unscreened plasma
frequency we estimate
the total coupling  constant $\lambda_{tot}$ for the electron-boson
interaction
within the framework of the Eliashberg-theory adopting an effective single
band approximation. For LaFeAsO$_{0.9}$F$_{0.1}$ a weak to intermediately
strong coupling regime and a quasi-clean limit behaviour are found.
For a pronounced multiband case we obtain a constraint for various
intraband coupling constants which in principle allows for a 
sizable strong coupling in bands with either slow electrons or holes.

\end{abstract}
\pacs{1315, 9440T}
\submitto{\NJP}
\maketitle
\normalsize

\section*{\bf Content}

{\bf 1. \quad Introduction \hfill 2\\
2. \quad Reflectivity and its analysis \hfill 4\\
2.1. \ Studies on polycrystalline samples \hfill 4\\
2.2. \ BaFe$_2$As$_2$ single crystal studies \hfill 7\\
3. \quad FPLO analysis and comparison of unscreened plasma 
frequencies \hfill 10\\
4. \quad Discussion of possible related microscopic many-body 
effects \hfill 12\\
4.1. \ The Mott-Hubbard "bad metal" scenario \hfill 12\\
4.2. \ The weakly correlated As-polarization scenario \hfill 15\\
5. \quad The {\it el-boson} coupling strength and constraints for pairing mecha-\\ 
\indent \hspace{0.2cm} nisms \hfill 20\\
5.1 \hspace{0.2cm} One-band Eliashberg analysis \hfill 20\\
5.1.1 Comparison with other pnictide superconductors \hfill 23\\
5.1.2 Comparison with other exotic superconductors \hfill 25\\
5.2 \hspace{0.2cm} Multiband aspects \hfill 28 \\
6.  \hspace{0.3cm} Conclusion \hfill 29 \\
Acknowledgments \hfill 30\\
References} \hfill 30\\

\section{\bf Introduction } 

Naturally, shortly after the discovery \cite{kamihara}
of superconductivity  at  
relatively high transition temperatures $T_c$ up to 56~K in 
several iron
pnictide compounds  despite all efforts world wide, 
the underlying pairing 
mechanism and many basic physical properties both in the
superconducting and in the normal state are still debated. 
This concerns also the nonsuperconducting parent compounds, which often
show magnetic phase transitions. In such a situation a combined theoretical 
and experimental study of selected systems as well as a detailed
comparison of various members of the fastly growing family among themselves 
and also with other more or less exotic superconductors, especially
with the high temperature superconducting cuprates (HTSC) might provide  
useful insight into related problems. In this context the
unscreened plasma frequency $\Omega_{p}$ determined from optical 
measurements 
or electron energy-loss spectroscopy (EELS) at frequencies near
the visible range
\begin{equation}
\Omega_{p}=\frac{4\pi e^2n}{m^*}
\end{equation}
is a physical quantity of central interest in the present paper for 
several 
reasons. (i) It provides direct insight in the dynamics of free charge
carriers that are responsible for 
the superconductivity. In particular, it reflects also 
the 
possible influence of many-body effects in changing their effective
mass $m^*$ and measures the density of paired electrons (condensate 
or superfluid density), if as in the BCS-case
 $n=n_s$ holds. (ii) According to the 
famous Uemura plot
\cite{uemura} for $T_c$ vs.\ $n_s/m^*$, simple linear relations between 
the two quantities may be expected for many exotic type-II
superconductors, and $\Omega^2_{pl}$ would then be a direct measure of $T_c$. 
Such plots have been presented
 both for underdoped HTSC and for the pnictide
superconductors as well 
\cite{uemura,goko,carlo,luetkens,khasanov,pratt,hiraishi}. In some 
cases, especially for members of the 122-family, i.e.\ for Ba(Sr)Fe$_2$As$_2$
derived superconductors its applicability has been 
put in question 
\cite{hiraishi,ren}. Anyhow, since a Bose condensation of bipolarons is 
naturally 
directly 
related to the superfluid density \cite{alexandrov}, bipolaronic scenarios 
such as 
the recently proposed one by 
G.A. Sawatzky {\it et al.} \cite{sawatzky,berciu}  
are of special interest.  

The unscreened  plasma frequency is also of considerable
interest for the analysis of resistivity ($\rho(T)$)
data analyzed within the frequently used
Drude model. In particular, the high-temperature
linear $T$-region is often used to extract the transport electron-phonon
({\it el-ph})
coupling constant $\lambda_{tr}$ \cite{bhoi} using the simple relation
\begin{equation}
\lambda_{tr}=\frac{\hbar\Omega^2_{p}}{8\pi^2\mbox{k}_B}\frac{d\rho}{dT}.
\end{equation}
For standard superconductors $\lambda_{tr}$ does not differ much from 
the Fermi surface averaged {\it el-ph} coupling constant 
$\lambda_{el-ph}$ which 
governs the  transition temperature $T_c$ of the superconducting state and it 
is also responsible for the renormalization of the electronic specific heat
of a Fermi liquid
at low-temperature. 
Using our previous very first estimate of $\Omega_{p}$ 
for LaO$_{0.9}$F$_{0.1}$FeAs \cite{drechslercond},
equation (2) has been used \cite{bhoi,sadovskii} to consider a 
classical {\it el-ph} scenario with $\lambda \sim $1.3 
for the high $T_c$-values and 
for a high-temperature saturation-like behaviour 
of the resistivity in the Pr based 1111-iron pnictides. However,  
the present improved knowledge and data
do not
provide support for such a point of view 
(see below). For more realistic estimates within a two-band model approach see
reference \cite{kulic}.

Finally, $\Omega_{p}$ and $\lambda$ determine also 
the pair-breaking parameter
$\beta =\Omega^2_{p}\rho_0/8\pi(1+\lambda)T_{c0}$ for various unconventional
pairing states
 \cite{lin,petrovic,mackenzie,radtke}:
\begin{equation}
-\ln \left( \frac{T_c}{T_{c0}} \right)=\psi \left( \frac{1}{2}+
\frac{\beta T_{c0}}{2\pi T_c} \right) -\psi \left( \frac{1}{2}\right), 
\end{equation}
where $\psi (x)$ is the digamma function and $T_{c0}$ denotes
the transition temperature being at maximum in the clean-limit
where the residual resistivity $\rho_0 \rightarrow 0$.

Here we report on the optical analysis of superconducting and 
nonsuperconducting parent compounds. Besides our own data also published data 
of other groups
will be discussed and used in our analysis.
We briefly introduce the employed codes for the determination of the
electronic structure and the unscreened plasma frequencies, which we regard 
as quantities not affected by several possible many-body effects
such as correlation, nonadiabatic, polaronic and polarization effects 
being under debate at present. With the aid of penetration depth data near
$T=0$ taken from $\mu$SR (muon spin rotation (relaxation))-data we 
are in a position to estimate the total coupling 
strength 
between the charge carriers (both
electrons and holes) and some still unspecified bosonic mode(s).
The analysis of the background dielectric constant 
$\varepsilon_{\infty}$ in the range of the screened plasma frequency 
as derived from optical measurements
sheds light
onto the role played by correlation and polarization 
effects which is a central 
problem of general 
interest in modern solid state physics independently on the peculiar 
physics of high-temperature superconductivity.

\section{\bf Reflectivity and its analysis}
\subsection{Studies on polycrystalline {\rm LaO$_{1-x}$F$_{x}$FeAs} samples}
Polycrystalline samples of LaO$_{0.9}$F$_{0.1}$FeAs \cite{drechsler,fuchs1} 
as well as of the undoped parent system LaOFeAs (also called (1111)
compound) have been 
prepared as described in previous work \cite{luetkens,zhu}. 
In general, the situation here parallels that for the HTSC. Right
after their discovery, also only polycrystalline samples were available, 
and have been studied by reflectance measurements. Further, also the
HTSC are quasi-2D metals with the highest plasma energy (PE) value for 
light polarization within the (a,b) crystal plane. In the case of HTSC,
there is a number of publications devoted to
 polycrystalline pellets, exactly as in our studies of 
La(O,F)FeAs, which demonstrate that the reflectivity
onset seen as a kink in the measured curves is a good measure of the 
in-plane PE value (see e.g.\  Kamaras K {\it et al.} \cite{kamaras}). Such
studies already provided very important information on the optical 
properties of HTSC, e.g.\ they were used to study the doping dependence of
the in-plane PE, and most importantly these results later on were 
corroborated by single crystal studies (see e.g.\ reference \cite{basov}).

For the optical measurements reported here the pellets have been polished to 
obtain appropriate surfaces. The reflectance measurements have been
performed using a combination of Bruker IFS113v/IFS88 spectrometers. This 
allows us to determine the reflectivity from the far infrared up to
the visible spectral region in an energy range from 0.009 up to 3~eV. The 
measurements have been carried out with different spectral resolutions
depending on the energy range, these vary form 0.06
 to 2~meV. Since the observed spectral features are significantly broader 
than these values, this slight variation in resolution does not impact our
data analysis. All reflectance measurements were performed 
at $T=$~300~K.

\begin{figure}[t]
\hspace{3cm}
\includegraphics[width=12cm]{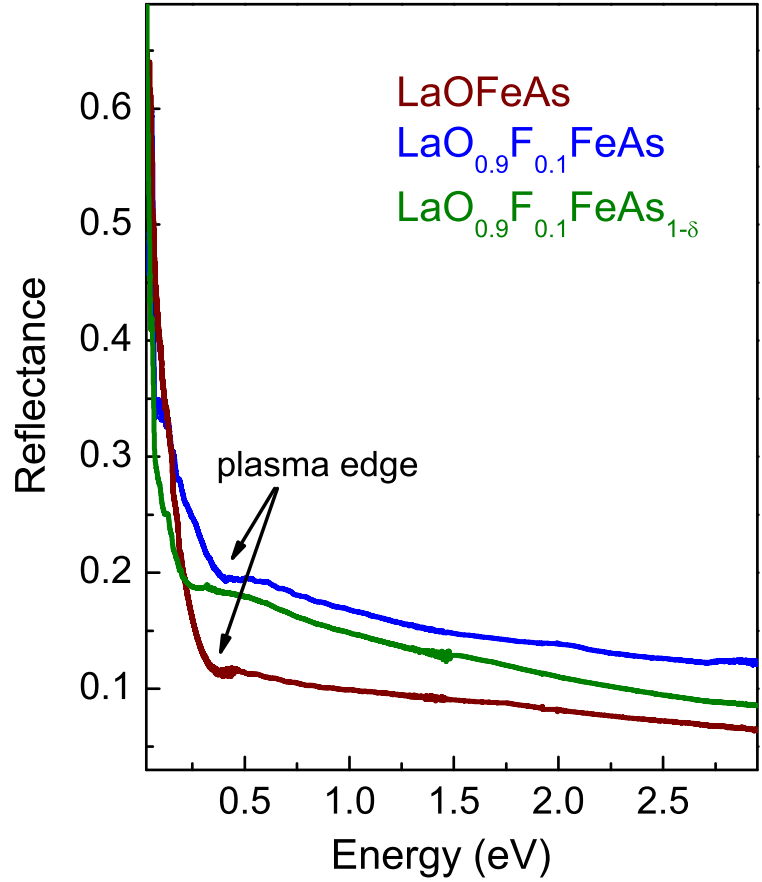}
\hspace{-2cm}
\caption{Reflectance of polished undoped
LaOFeAs, optimally doped LaO$_{0.9}$F$_{0.1}$FeAs, as well as an 
arsenic-deficient sample
LaO$_{0.9}$F$_{0.1}$FeAs$_{1-\delta}$ with $\delta \approx 0.1$. 
All three curves are characterized by 
a rather pronounced kink which is
attributed to a plasma edge at about 380~meV and 395~meV,  respectively. 
For the interpretation of the broad edge-like feature near 240~meV
in the special case of LaO$_{0.9}$F$_{0.1}$FeAs$_{1-\delta}$ see figure~2 
and text.
In addition, there are signatures of interband excitations around 0.6~eV
(and for LaO$_{0.9}$F$_{0.1}$FeAs at about 2~eV).} \label{reflectivity}
\end{figure}

In figure~1 we show the reflectance of our LaO$_{1-x}$F$_{x}$FeAs 
samples 
($x=0, 0.1$). The reflectance significantly drops between 0.13 and
0.4~eV, thereafter it shows a small increase until about 0.6~eV 
before it 
starts to slightly decrease again. In addition, at about 2~eV another
small upturn is visible for LaO$_{0.9}$F$_{0.1}$FeAs. We attribute the two 
small features at about 0.6 and 2~eV to weak electronic interband
transition at the corresponding energies. This observation is in accord 
with the behavior of the optical conductivity $\sigma(\omega)$ predicted
by Haule {\it et al.} \cite{haule} using dynamical mean-field theory (DMFT), 
where As 4$p$ to Fe 3$d$ interband transition slightly below 2~eV and a weak
feature  near 0.6~eV for the undoped parent compound LaOFeAs have been 
found. Again, our measurements on powder samples do not allow the
extraction of their polarization. The steepest edge centered at about 
250~meV with an high energy onset at about 380(20)~meV for LaOFeAs and
395(20)~meV for LaO$_{0.9}$F$_{0.1}$FeAs represents the plasma edge or 
plasma energy, which is observed at relatively low values. In
consideration of the quasi-2D character of the electronic states in 
LaO$_{1-x}$F$_{x}$FeAs and the expected strong anisotropy of the plasma
energy for light polarized $\parallel$ to the ({\bf a,b}) and {\bf c} 
crystal axes, respectively, the onset of the plasma edge in figure~1 gives a
reasonable value of the in-plane value (i.e.~for light polarized within 
the ({\bf a,b}) crystal plane). Consequently, the
in-plane plasma energies for the two LaO$_{1-x}$F$_{x}$FeAs samples 
(undoped and 10\% electron doped) is almost independent of the doping 
levels studied here.

\begin{figure}[t]
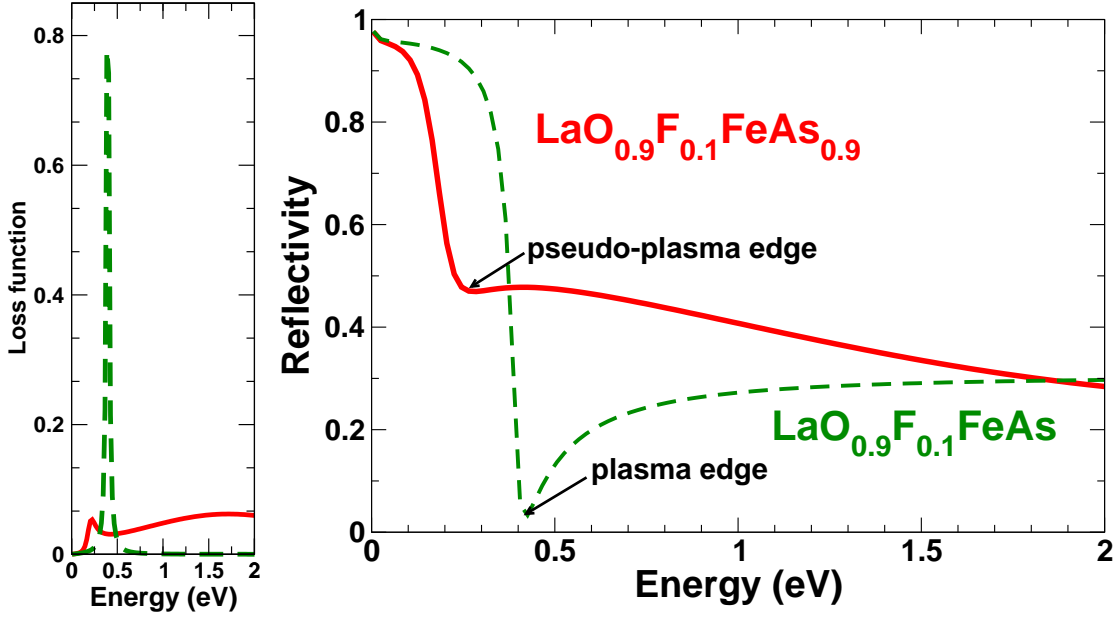

\vspace{2cm}
\begin{minipage}{3.5cm}
\includegraphics[width=3.4cm]{fig2a.eps}
\end{minipage}
\begin{minipage}{11.2cm}
\includegraphics[width=11.1cm]{fig2b.eps}
\end{minipage}
\caption{Idealized in-plane Drude-Lorentz model for the 
loss function (left) and the reflectivity (right) 
of optimally doped LaO$_{0.9}$F$_{0.1}$FeAs, as well as an 
arsenic-deficient sample
LaO$_{0.9}$F$_{0.1}$FeAs$_{1-\delta}$ with $\delta \approx 0.1$. 
For a qualitative comparison with experimental polycrystalline data see 
figure 1. For fit parameters see text. 
} \label{f2r}
\end{figure}

Within a simple Drude model, the value for the plasma energy allows a first,
 rather crude estimate of the charge carrier density in
LaO$_{1-x}$F$_{x}$FeAs. Within this model, the  observed screened
plasma energy, $\omega_p$, is given by 
$\omega_p^2 =4\pi ne^2/m \varepsilon_\infty$,
with $n$, $e$, $m$, and $\varepsilon_\infty$ representing the charge 
carrier density, the elementary charge, the effective mass of the charge
carriers and the background dielectric 
screening due to higher lying 
electronic excitations \cite{wooten}. Adopting an 
effective mass equal to the
free electron mass, and $\varepsilon_\infty \sim 12$ \cite{boris}, one 
would arrive at a rather low charge density of about $9\cdot 10^{-20}$
cm$^{-3}$. We note, that this estimate ignores the strongly anisotropic 
nature of the electronic orbitals near the Fermi energy in
LaO$_{1-x}$F$_{x}$FeAs. A more detailed discussion of the 
unscreened in-plane 
plasma 
energies found well above 1.3~eV but below 2~eV  
in all pnictide superconductors is given in the next sections
see also table~2.

Finally, we examined also an arsenic-deficient sample
LaO$_{0.9}$F$_{0.1}$FeAs$_{1-\delta}$ with $\delta \approx 0.1$
\cite{fuchs1,fuchs2}. Here at first glance a plasma-edge like feature
near $\omega_p=0.24$ eV, only,
would suggest that
the screened plasma frequency is considerably lowered.
(see figure 1). Adopting a linear scaling
of the background polarizability $\varepsilon_{\infty}-1$ with the 
As-concentration, one arrives at $\varepsilon_{\infty}\approx 10.5$ 
instead of 12 for the non-deficient sample. 
Within a nonlinear scaling suggested by the Clausius-Mossotti
relationship between the dielectric constant and the ionic polarizibilies
discussed in more detail in section 4.2 one estimates even a slightly
smaller value of $\varepsilon_{\infty}\approx 9.97$.  
Thus, one would arrive   
at $\Omega_{p}=0.79$ to 0.76~eV for the unscreened plasma frequency. 
However, 
such a strong change of $\Omega_{p}$ by about 0.6 eV (!) would 
imply very large changes in the 
electronic structure (e.g.\ by filling the 
hole pockets and an additional mass enhancement) or 
alternatively a twice as large
$\varepsilon_{\infty}$. Both scenarios are difficult to imagine 
microscopically. For this reason we will consider a third possible
explanation with  
an almost unchanged $\Omega_{p}$ and a slightly reduced  
$\varepsilon_{\infty}\approx 10$. We ascribe the clearly
changed optical properties to the presence of arsenic derived electrons
and adopt that an absorption peak described by a
 Lorentzian curve like that absorption near 0.66 eV in the BaFe$_2$As$_2$
single crystal data  with about 5\% arsenic vacancies
\cite{ni} (see below) might be mainly responsible for that unexpected 
feature. Simulating qualitatively the polycrystalline reflectance data
shown in figure 1
by an effective Drude-Lorentz model we arrive at a corresponding 
Lorentzian curve peaked near 1.16~eV
and a slightly lowered unscreened plasma frequency $\Omega_p=$1.3~eV
(A quantitative fit would require a more or less realistic account
of the shapes of the microcrystalline grains and their anisotropic dielectric
properties which, however, is without the scope of the present paper.). 
The oscillator strength of this Lorentzian-peak 
is expected to be about twice as large which is 
supported by our fit value of 6.5 eV, see figures 2-4 and compare table 1.
The different binding energy might be related to the rather different 
As-environment in the 1111 and the 122 compounds with no La(O,F) layer 
in the latter and an enhanced pinning provided by the La$^{3+}$ ions being
close to the As-sites whereas in the former case it is provided by the 
less effective pinning due to Ba$^{2+}$-ions.
The resulting loss function and the optical conductivity 
are shown in
figure \ref{loss-sigma}. The obtained optical conductivity resembles
the mid-infrared absorption observed by Yang {\it et al.} in a 
Ba$_{0.55}$K$_{0.45}$Fe$_2$As$_{1-\delta}$ single crystal \cite{yang,ni}
with a $T_c\approx 30$~K slightly above our arsenic-deficient La-111 sample.
This observation gives further support for our vacancy derived 
bound-electron 
scenario. 
In this context it would be 
interesting to study analogous effects in the related 
superconducting FeSe$_{0.88}$ and other substoichiometric iron telluride
and selenide
compounds \cite{lee}.

Here we would like to mention that a very small value 
for the unscreened plasma frequency has been derived from an ellipsometry
study of LaO$_{0.9}$F$_{0.1}$FeAs and a subsequent analysis applying 
the effective medium approximation (EMA) to obtain 
$\sigma_{\mbox{\tiny eff}}(\omega)$ for a polycrystalline 
sample \cite{boris}. For 
spherical grains and $\sigma_{ab}\gg \sigma_c$ as expected for a 
quasi-2D metal
(as is the case for LaO$_{0.9}$F$_{0.1}$FeAs), the EMA predicts 
$\sigma_{\mbox{\tiny eff}}\approx 0.5\sigma_{ab}$, which then would 
correspond
to a slightly larger value of $\tilde{\Omega}^{ab}_p$~=~0.86~eV.
In addition some 
spectral weight from the Drude part was lost in reference \cite{boris}
due to the
fitting of broad Lorentzian-peaks also for the   
two low-$\omega$ interband transitions $\alpha_1$ and $\alpha_2$
at 0.75 and 1.15~eV, respectively. Adding this lost weight to the Drude 
term we arrive at an enhanced corrected Drude weight by a factor of 1.28
which corresponds to a corrected
$\tilde{\Omega}^{ab}_p \approx $0.975~eV. 
Moreover, since this value has been derived mainly from the 
low-$\omega$, damped
Drude region (below 25~meV), it is still renormalized by 
electron-boson coupling, and a high-energy boson well 
above 25~meV has been assumed to
explain the high $T_c$ at weak coupling. Using a weak 
total {\it electron-boson} coupling 
constant $\lambda\approx 0.61$ ( see section  5.1) one 
estimates
$ \Omega_p=$1.24 eV, close to a value of 1.37 eV suggested from 
the plasma energies in figure~1 and dielectric screening (see above).

\subsection{BaFe$_2$As$_2$ single crystal studies}

With the successful synthesis of single crystals, 
more detailed investigations of the optical response of iron pnictide 
superconductors become
possible. In particular, the background dielectric 
constant $\varepsilon_{\infty}$ can be
determined more safely. 
Reflectivity data of Ba$_{1-x}$K$_x$Fe$_2$As$_2$ and SrFe$_2$As$_2$ single 
crystals have been reported recently, and their analysis
corroborates what has been derived from the polycrystalline samples as 
discussed above. In figure~3 we present reflectivity data of a
BaFe$_2$As$_2$ single crystal and a light polarization within the 
(a,b) crystal plane.
The investigated single crystal with the dimension of about 
0.5x0.5x0.1 mm was grown from a
tin flux.
Elemental Ba, Fe, As were
added to Sn in the ratio 1 : 2 : 2 : 15 and placed in a corundum 
crucible
which was sealed in an argon filled silica ampoule. The sealed ampoule 
was
heated in steps to 850$^\circ$C with a rate of 20~K/h, hold for 
36~hours and cooled down. The tin flux was removed by solving in 
diluted HCl. The resulting
gray, crystals with metallic luster were washed 
in distilled water 
several
times and finally dried.

The studied single crystal shows a
thin platelet-like shape and 
a shiny (0, 0, 1) surface.
The thorough characterizations via specific heat, resistivity, magnetic 
susceptibility and
chemical composition analysis have been carried out on the crystals 
obtained 
from the same
batch. Similar physical properties to those in the crystals reported in 
references \cite{ni} have been
obtained. The EDX composition analysis of several specimens of this 
batch resulted an
average composition of Ba$_{0.95}$Sn$_{0.05}$Fe$_2$As$_2$ and we expect 
$\sim~5$\% Sn to be incorporated in the
crystal structure. 
For the consequences of an As-deficiency with respect to 
low-frequency optical properties see below and for 
the upper critical field $B_{c2}(T)$ of the K-doped system 
see \cite{fuchs2}.

\begin{figure}[t]
\hspace{3cm} \includegraphics[width=12cm]{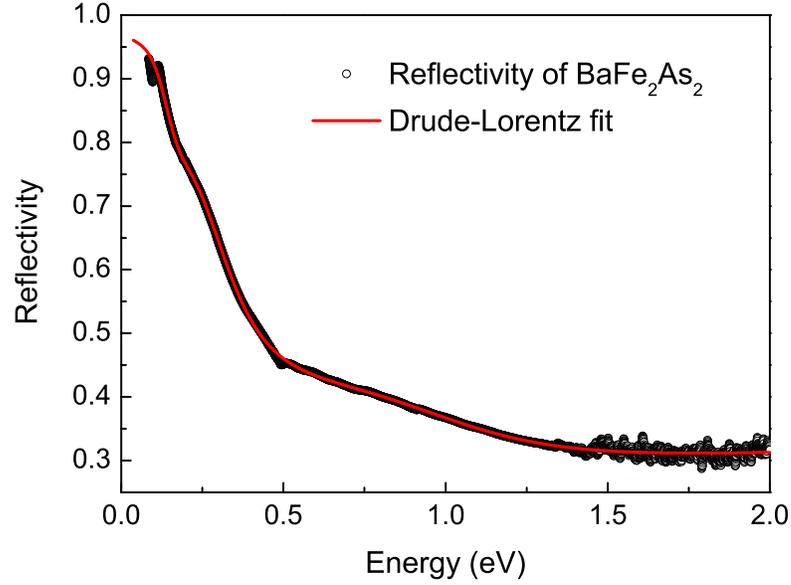}
\caption{Reflectivity of a BaFe$_2$As$_2$ single crystal and a light polarization within the (a,b) crystal plane (open circles). The solid line
represents the result of our Drude Lorentz analysis (see text).} \label{f3}
\end{figure}
\begin{figure}[b]
\hspace{0.6cm}
\begin{minipage}{7.0cm}
\includegraphics[width=6.9cm,angle=-90]{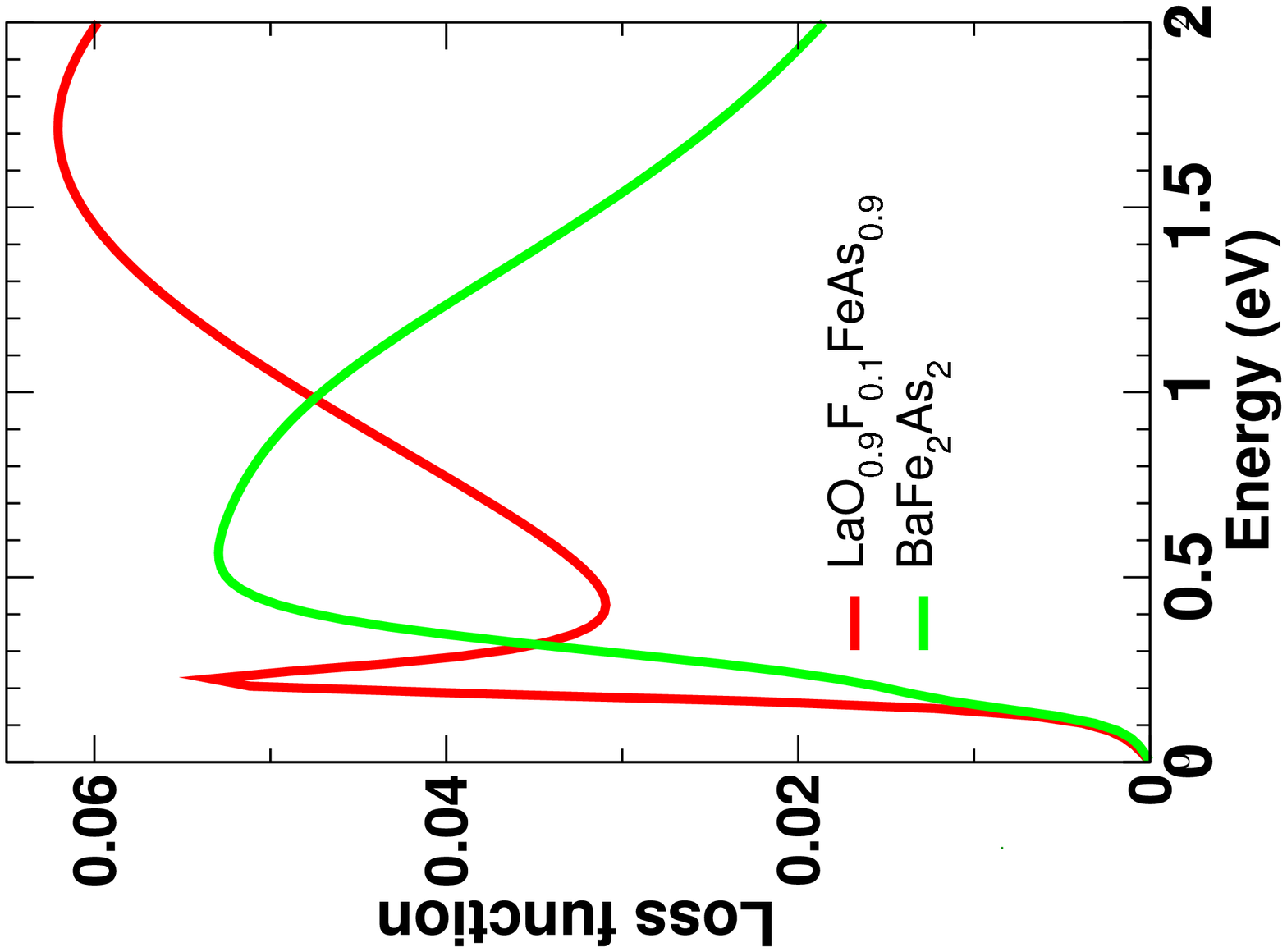}
\end{minipage}
\begin{minipage}{7.6cm}
\includegraphics[width=7.cm,angle=0]{fig4b.eps}
\end{minipage}
\caption{Experimental in-plane loss function (left) 
and optical conductivity (right) of a BaFe$_2$As$_2$ single crystal 
and the same 
for the suggested minimum 
Drude-Lorentz model
for the 
arsenic-deficient La-1111 system.
Adopting $\omega_{0,1}=$1.16 eV $\gamma_{0,1}=$ 3.3 eV 
and $\Omega_{p,1}$=6.5 eV for 
the "bound" electrons as well as
$\Gamma_p=$0.03 eV and 
$\Omega_p=$1.3 eV for the free electrons for the latter.
Compare table \ref{t1} and equation (\ref{G1}).}
 \label{loss-sigma}
\end{figure}
Also these data are characterized by a "kink"-like feature
(but not by a sharp minimum as expected in the
 case of a clean single crystal!)
 in the reflectivity 
around 0.4 eV, with a steeper decrease at lower energy values. The
polycrystalline samples as analyzed in the previous section did not 
allow a quantitative analysis of the reflectivity values due to 
relatively
rough surfaces, but in the case of single crystal data as shown in 
figure~2 this is feasible, and figure~2 also shows a fit of our 
reflectivity data
within a Drude-Lorentz model:
\begin{equation}
\epsilon(\omega) =\epsilon_{\infty}-  \frac{\Omega_{p}^2} {\omega^2+i\omega\Gamma_p}  +\sum_{j=1}^{3}\frac{\omega_{p,j}^2}
{\omega_{0,j}^2-\omega^2-i\omega\gamma_j} \ . 
\label{G1}
\end{equation}
This fit describes the data very well and allows the extraction of more
 physical information. The resulting fit parameters are summarized in
table~1.

\begin{table}
\caption{Parameters derived from a Drude-Lorentz fit of the reflectivity of a 
BaFe$_2$As$_2$ single crystal. Here the following notation has been used:
$\Omega_{p}$
= Drude plasma frequency, $\Gamma$ = Drude width, $\omega_{0,j}
$ = Lorentz oscillator position, $\gamma_j$ = Lorentz oscillator width,
$\omega_{p,j}$ = individual Lorentz plasma frequency (
or oscillator strength), and $\epsilon_{\infty}$ =
 background dielectric constant).}
\vspace{0.5cm}
\hspace{2.8cm}
\begin{tabular}{c|ccc||ccc}
\hline
j & $\omega_{0,j} (eV)$  & $\gamma_j$ (eV) 
& $\omega_{p,j} (eV)$ & $\Gamma_p$ (eV) & $\Omega_{p} (eV)$ & 
$\epsilon_{\infty}$\\
\hline
1 & 0.2 & 0.16 & 0.79 & 0.031 & 1.58 & 14.9\\
2 & 0.66 & 2.9 & 4.31 &0.087 \cite{hu} &1.6 \cite{hu}&\\
3 & 0.93 & 1.63 & 1.86 \\
\hline
\end{tabular}

\label{t1}
\end{table}

One important finding of our fit is the value for the screened 
plasma frequency $\omega_{p}=\Omega_p/\sqrt{\epsilon_{\infty}}$ 
of about 400 meV, 
which is in
very good correspondence to what has been derived from the polycrystalline 
LaO$_{1-x}$F$_{x}$FeAs samples above. This demonstrates that this
value is rather independent of the stoichiometry but 
is a 
characteristic feature of the FeAs planes common to all FeAs based
superconductors. Furthermore, in order to obtain a reasonable description 
of our data within the Drude-Lorentz approach, it is necessary to
include Lorentz oscillators above and {\it below} this unscreened plasma 
energy 
(see table 1) at least for the BaFe$_2$As$_2$ (also called (122)
compound), and other (122) systems. 

Regarding the clean (1111) systems, 
so far clearly visible interband transitions have been observed only above 
the
plasma edge \cite{drechsler,boris}. 
An inspection of the atomic occupancies given for such single crystals 
obtained from
the Sn-flux by \cite{ni}, reveals a non-neglible amount of disorder, 
first of all
of arsenic-vacancies about half as large as in the polycrystalline 
arsenic-deficient
sample reported above. For this reason we assume that the low-frequency 
Lorentzian-peaks 
describe in both cases transitions from As-vacancy introduced,  
more or less localized, states
below the Fermi-level $E_F$ 
but well above the usual As $4p$ states which occur
about 2~eV below $E_F$. These states are thought to accomodate the 
large amount of electrons compared for instance with its free electron Hall
number $n_H\approx 5\cdot 10^{20}$cm$^{-3}$ \cite{fuchs2}: 
$ \approx c_{vac}Z/\tilde{v}_{\mbox{\tiny cell}}\approx 10^{21}$cm$^{-3}$
which would otherwise heavily overdope the La-1111 system or electron dope 
the nominal 122 parent 
compound, where $Z <$3 is the effective anionic As-charge, 
$c_{vac} \approx $ 0.05 to 0.1 denotes 
the atomic As-vacancy
concentration an $v_{\mbox{\tiny cell}}=$70.75\AA$^3$ 
denotes the molar cell volume.

Although the reflectivity curves 
for FeAs based single crystals as presented in figure~3 and in the literature
so far \cite{yang,hu,li,Li,pfuner,quazilbash} do not agree in all aspects, there are 
essential 
parameters in common, which give quite a lot of insight into the
physics of these systems. First, interband excitations are present in the 
energy ranges around 0.2-0.3 eV and 0.6 - 2 eV which most likely arise
from the manifold of iron related bands that cross the Fermi 
level \cite{singh}. Second, the background dielectric screening in the 
energy range
of the plasma energy is quite large with values of about 
12 to 16.
Noteworthy, comparable values have been already observed occasionally
also for other transition metal compounds  \cite{nava}. 
Third, the value of the 
(unscreened) plasma frequency 
is in
the range of 1.4 - 2 eV, rather independent of the doping level, and 
importantly, it is substantially smaller than what is predicted from 
density functional (DFT) based band
structure calculations. The latter issue will be discussed in detail in 
the following sections.

\section{\bf FPLO analysis of the unscreened plasma frequency}
The theoretical plasma frequencies have been taken 
from DFT based 
calculations in the local (spin) density approximation (L(S)DA). 
We used the full 
potential
local orbital code \cite{FPLO} in version 7 with the standard 
double numerical
basis. The theoretical unscreened (Drude) 
plasma frequencies are derived from the Lindhard intra band
expression $\Omega_{p}$ 
(see references \cite{cohen,mattheiss} for the isotropic 
case and equation (\ref{plasma-LDA}) below for 
the straightforward
anisotropic generalization). They  
basically represent the Fermi surface average of the Fermi
velocity squared. 

It has been checked that the results are converged 
within
a few percent with respect to the $k$-integration scheme.
We used the experimental lattice constants and the theoretical 
arsenic position.
However, the frequencies seem not to be too sensitive to the lattice
parameters. In the case of doped systems the 
simple virtual
crystal approximation has been used. 
Thus, here we report calculations of the eigenvalues of the anisotropic
conductivity (dielectric function) tensor
\begin{equation}
\hbar^2\Omega^2_{pl,\alpha \alpha}=4\pi e^2N(E_F)v^2_\alpha, 
\quad \alpha= x,y,z
\label{plasma-LDA}
\end{equation}
where the Fermi velocity components $v_\alpha$
have been averaged over all Fermi
surface sheets (FSS) and $N(E_F)$ denotes
 the total density of states at the Fermi level $E_F$
for both spin directions.
In the tetragonal phase the '$xx$' and '$yy$' component do coincide
and we will them denote as in-plane plasma '$ab$'-frequency whereas the
out-of plane '$c$'-plasma frequency is given by the '$zz$' component.
In terms of individual Fermi surface sheets (bands) equation 
(\ref{plasma-LDA}) can be
rewritten as
\begin{equation}
\Omega^2_{p,tot}=\sum_i\Omega^2_{p,i}
\label{multiplasmon}
\end{equation}
where $i$ denotes the band index.
In the pnictides under consideration all DFT-calculations
predict up to five bands (FSS) at $E_F$ consisting of electron ($el$)
and hole ($h$) FSS. For the sake of simplicity the 
discussion of multiband effects all these FSS can be lumped into 
one effective $el$-FSS and one $h$-FSS, thus resulting in frequently 
used two-band models. 

Due to the layered structure of the pnictides one
has $\Omega^{ab}_{p} \gg \Omega^c_{p}$.
The larger in-plane distance and the smaller transfer integrals in
$c$-direction for the (1111) systems as compared for instance with the
(122) systems is mainly responsible for the relative small
values of both in-plane and out-of plane comnponents.
For comparison we included also calculations by Singh \cite{singhprivate}
for the case LaOFeP
and Boeri {\it et al.} \cite{boeri} for the case of LaOFeAs
and LaOFeP where different bandstructure codes have been used.
We note that different codes results in essentially the same
unscreened plasma frequency. All are clearly a factor of 1.5 to 2
larger
compared with the experimental values
(see table 2),
pointing to the presence
of many-body effects which for instance may cause a mass renormalization.
Several possibilities to resolve this issue will be discussed below.

\begin{table}
\caption{\label{math-tab2}Comparison of calculated (DFT)
and experimental unscreened
in-plane $(ab)$
plasma frequencies measured in eV for polycrystalline samples (PC) and single
crystals (SC). For completeness, the theoretical 
out-of plane values are shown,
too. In the references the first citations notation denotes the experimental 
optical data (if available) and the second one the structural data
employed in the DFT calculations. }
\vspace{0.3cm}

\begin{tabular*}{\textwidth}{@{}l*{19}{@{\extracolsep{0pt plus
12pt}}l}}
\br
Compound&DFT (ab)& DFT (c) &Experiment (ab)&Sample&Reference\\
\mr
LaOFeAs  &2.1 - 2.3&0.25-0.34& 1.33  &PC& present work\\
LaO$_{0.9}$F$_{0.1}$FeAs  &2.1 - 2.3&0.25-0.34& 1.36  &PC&\\
LaO$_{0.9}$F$_{0.1}$FeAs  &2.1 - 2.3&0.25-0.34& 1.37  &PC&\cite{drechsler}\\
\qquad"&&&0.6 (1.24) &PC&\cite{boris} see text\\
LaO$_{0.1}$F$_{0.1}$FeAs$_{0.9}$  & & & $\leq$ 1.3  &PC& present work\\
SrFe$_2$As$_2$ &2.8 &1.36&1.72  &SC& \cite{hu}\\
La$_x$Sr$_{1-x}$Fe$_2$As$_2$&2.8&1.36 - 1.46&&&\\
 SrFe$_{2-x}$Co$_x$As$_2$&2.7&1.35&&\\
BaFe$_2$As$_2$&2.63&0.8&1.72  &SC&\\
\qquad " &&&1.6 &SC& \cite{hu} \\
\qquad " &&&1.58 &SC& present work \\
\qquad "&&&1.39 &SC& \cite{pfuner} \\
K$_x$Ba$_{1-x}$Fe$_2$As$_2$&2.63&0.8&1.5  & SC& \\
K$_{0.45}$Ba$_{0.55}$Fe$_2$As$_2$&2.63&1.72&1.9 $\pm$ 0.2 \quad &SC& 
\cite{yang}\\
CaFe$_2$As$_2$& 2.95&1.96&&\\
LiFeAs&2.9&0.83&&&\cite{tapp}\\
NaFeAs&2.68&1.16&&& \cite{parker}\\
LaOFeP&2.37&0.47&1.85&  SC& \cite{quazilbash}\\
\br
\end{tabular*}
\end{table}
The inspection of the calculated and empirically found unscreened plasma 
frequencies 
shows a very sensitive dependence 
 on the Fe-Fe-distance which in the tetragonal phase is given by the 
in-plane lattice constant $a/\sqrt{2}$. Since 
$\Omega_p \propto v_F \propto t_{dd}a$, 
where $t_{dd}$ is a representative value for 
the in-plane Fe-Fe 3$d$
transfer integrals, a simple power-law has been predicted 
considering the related Slater-Koster integrals
\cite{harrison}. As a result we expect
\begin{equation}
\Omega_p \approx \Omega_{p,0}/(a/a_0)^{\nu}, 
\label{plasma-scaling}
\end{equation}
where $a_0=$4 \AA \ has been introduced for convenience. According to 
Harrison \cite{harrison} we adopt $\nu = 4$ and arrive at 
at $\Omega_{p,0}=1.4$~eV. Applying the approximate expression 
(\ref{plasma-scaling}) to rare earth-1111 systems, where 
to the best of our knowledge
no high-frequency optical data are available, and considering
for instance two superconductors with the highest
transition temperatures $T_c > 50$~K: SmO$_{0.85}$FeAs and NdO$_{0.85}$FeAs 
with $a=3.897$~\AA \ and $a=3.943$~\AA \ 
we estimate $\Omega_p=1.55$~eV and 1.48~eV, respectively.  

Having now a large enough set of available empirical 
plasma frequencies $\Omega_p \sim 1.4$ to 1.7~eV we are in a position
to re-estimate the mentioned above transport coupling constants 
\cite{bhoi,sadovskii}. With these values using equation (2) we would
arrive, 
instead at $\lambda \sim \lambda_{tr}=1.3$, 
at a {\it superstrong} coupling regime $\lambda \sim 4$ to 5 (!).
 Then we should
 have to ask, why is $T_c$ of the iron pnictides so low? 
In our opinion other scattering 
mechanisms than the usual {\it el-ph} interaction, in the 
{\it el-el} or the {\it el-spin fluctuation} channels,
 contribute to $\rho(T)$ and possibly dominate it in 
the quasilinear region. Adopting a weak {\it el-ph} coupling 
constant $\sim $0.2 like the one estimated by Boeri {\it et al.} \cite{boeri}
a second  linear region might be achieved beyond 600 K, only. To the best 
of our knowledge there are no experimental data available at present 
for that region.

\section{\bf Discussion of possible related many-body effects}
\subsection{The Mott-Hubbard scenario}
Our LDA-FPLO
band structure calculations for the
nonmagnetic ground
state of LaOFeAs provide a value for the
plasma frequency within the ({\bf a,b}) crystal plane of
$\Omega^{LDA}_p
=2.1$~eV
and a small value for a polarization along
the {\bf c} axis of 0.34~eV in accord with
2.3~eV and 0.32~eV given
in reference \cite{boeri}. Note that these "bare" values do not
describe the screening through $\varepsilon_{\infty}$
caused by interband transitions.
In order  to get the measured 
values of 
$\omega_p$~=~0.395~eV for the fluorine doped and 0.38~eV for the undoped
parent system
an unusually large value of
$\varepsilon_{\infty} \sim $~28.3 to 33.9 would be required
which seems
to be unrealistic and instead intermediate values
$\varepsilon_{\infty} \sim 12$ to 15 
considerably larger than those for cuprates
(4 to 6) would be
 expected \cite{boris}. Then, for the empirical unscreened plasma 
energy $\Omega_p$
a value about
1.4~eV would be expected on the basis
of our reflectance data.
Thus the question arises, what is the origin of these renormalizations?
In view of the transition metal iron it is natural to consider
moderate short range Coulomb interaction induced 
correlations as a possible reason.
 To get more insight 
into the role of possible 
correlation effects adopting the parameter set of 
Haule {\it et al.} \cite{haule}
we calculated at first numerically
the interband contribution to the 
complex dielectric function $\varepsilon(\omega)$
from
the in-plane DMFT optical conductivity $\sigma(\omega)$ between 1 and 6~eV
for the undoped LaOFeAs
given in figure~5 of reference~\cite{haule}.
\begin{figure}[t]
\hspace{3cm}\includegraphics[width=12cm]{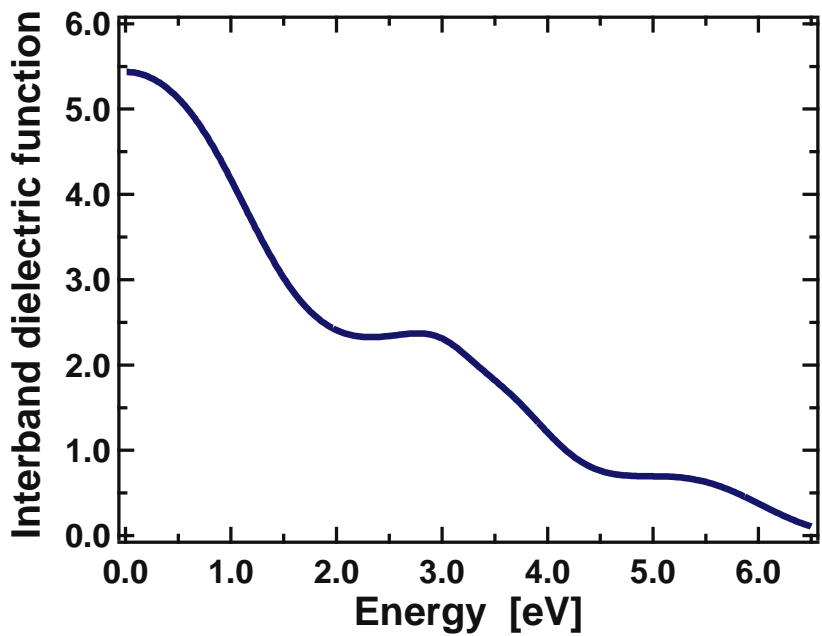}
\caption{Interband contribution to the real part of the dielectric function
$\varepsilon_1(\omega)$ of LaOFeAs
using the DMFT-calculation for its $\sigma (\omega)$ as given by 
Haule {\it et al.} in reference
\cite{haule} in figure~5 therein with $U_{\rm d}=$ 4 eV. } 
\label{fig5}
\end{figure}
\begin{figure}[b]

\begin{minipage}{7.8cm}
\includegraphics[width=7.7cm]{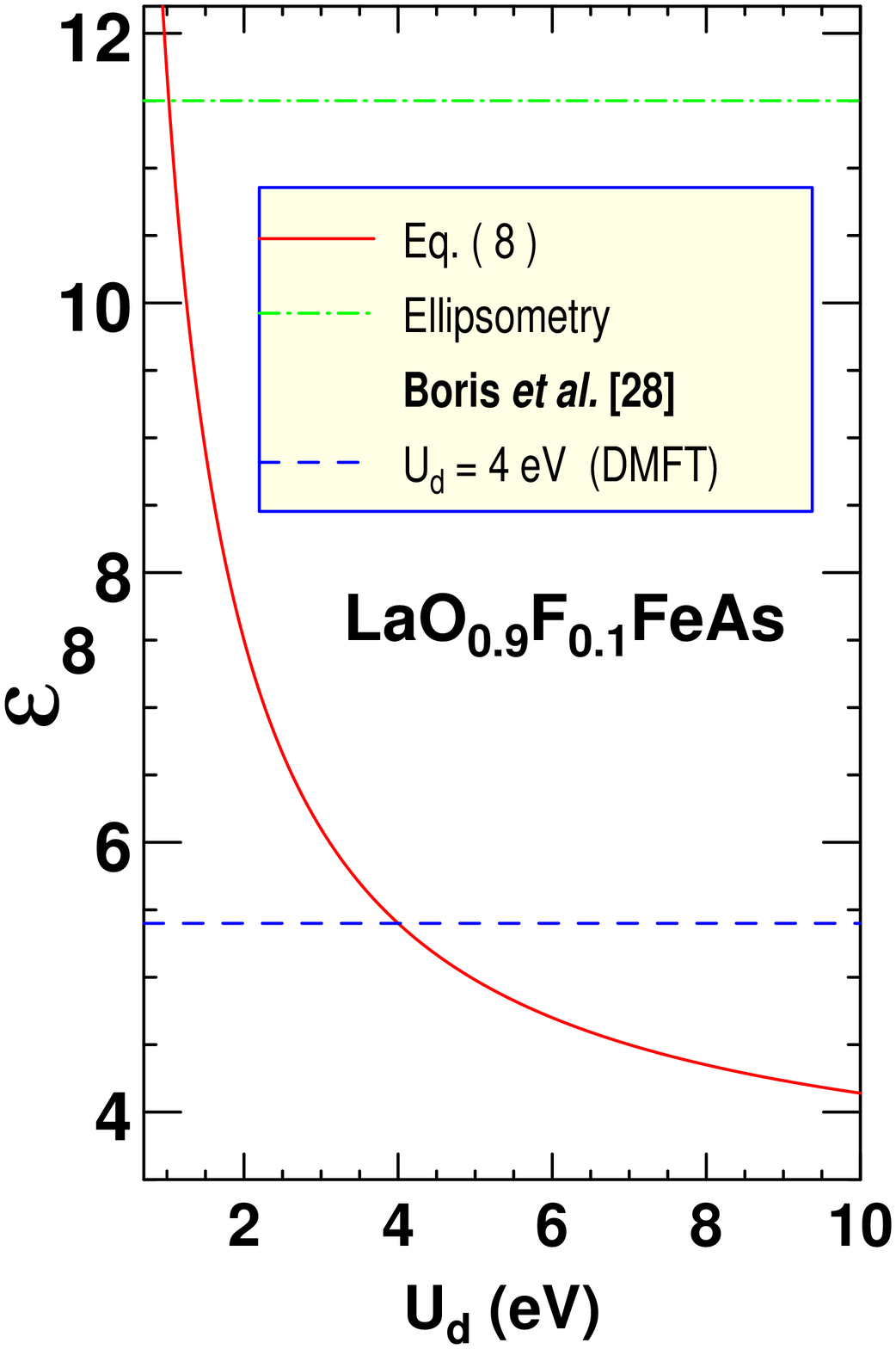}
\end{minipage}
\vspace{-0.5cm}
\begin{minipage}{7.8cm}
\includegraphics[width=6.cm]{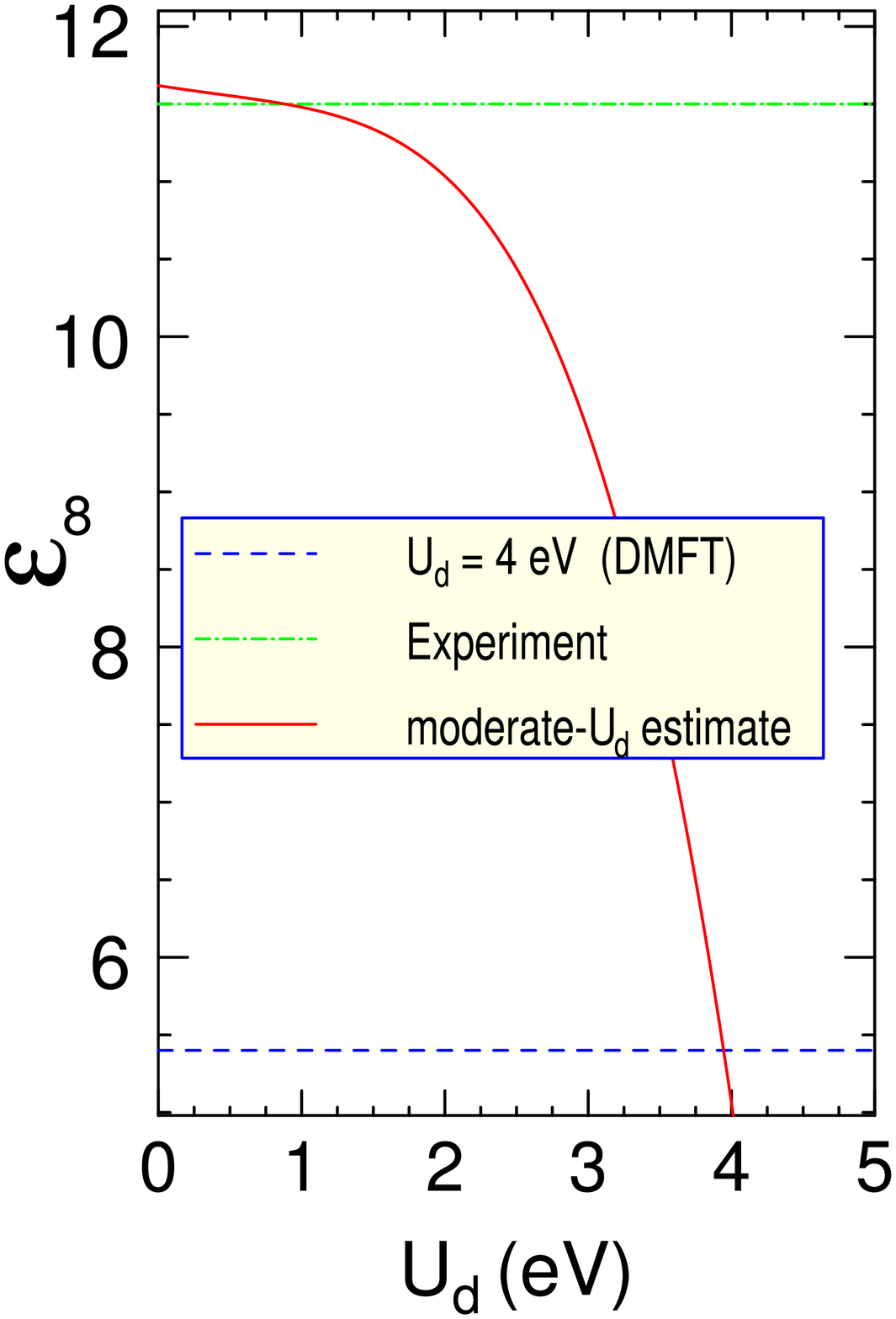}
\end{minipage}
\caption{Semi-empirical relation between
the total coupling constant from the mass enhancement
entering the penetration depth  $vs$.\ interband screening
modeled by the
dielectric background constant $\varepsilon_{\infty}$ using equation (8)(left)
and a first estimate proposed in reference \cite{drechsler} (right)
(see also figure 7 (right)).}
\label{epsinftu}
\end{figure}
From its static value we obtained $\varepsilon_{\infty}$~=~5.4, only
(see figure~5).
Taking into account the small differences in the optical spectra
between the undoped LaOFeAs and the optimally doped LaO$_{0.9}$F$_{0.1}$FeAs
as shown in sections 2 and 3,
we will not distinguish between their $\varepsilon_{\infty}$-values
in the following discussion. 
(Using the ionic polarizabilities of F$^{-1}$ and O$^{-2}$ given in 
section 4.2 and
 the unit cell volume of $v$=142.32 \AA$^3$ \cite{kamihara} for LaOFeAs,
 one estimates from equation (9) a slight change of its
$\varepsilon_{\infty}$ to 11.61 or 11.25 for the parameter sets of 
references \cite{shannon,remark-arsenic} and 
\cite{dalgarno,clavaguera,zhaox,vineis,handbook,remark}, respectively,
compared with 11.5 to 12 for LaO$_{0.9}$F$_{0.1}$FeAs \cite{boris}.) 
In the strongly correlated (narrow band limit) limit $U_d \gg W$, for a very qualitative 
discussion, 
$\varepsilon_{\infty}$
can be written approximately in the following way
\begin{equation}
\label{varepsilon-U}
\varepsilon_{\infty}=1+\frac{f_{pd}}{E_{pd}}+\frac{f_{dd}}{U_d}+ ... ,
\label{DMFT-optik}
\end{equation}
where $W\sim $2~eV denotes the total Fe 3$d$ bandwidth, $E_{pd}\sim $ 2eV is the oscillator strength for the arsenic-iron 
$4p$-Fe$3d$ interband transitions and $f_{dd}$ describes the transitions 
through 
the Mott-gap. From our numerical simulations for $U_d=$4 eV we estimate 
$f_{pd}=$ 4.6~eV and $f_{dd}=$8.4~eV. The resulting curve is shown in 
figure~\ref{DMFT-optik} (left). 

Despite the crude estimate of equation \ref{DMFT-optik}
in the weak $U$-limit,
in general for  smaller
values for the onsite Coulomb repulsion $U_{d}$
within the DMFT-approach would result in somewhat larger 
$\varepsilon_{\infty}$ values due to an energy shift of the
interband
transition from the lower Hubbard band. 
Hence, a corresponding systematic DMFT study
with sufficiently smaller 
$U_{d}$ and $J_{d}$-values would
be of interest. Alternatively to figure~6 (left) one might expect
that for small $U$-values there should be no visible 
influence of $U$ at all. Then a negative curvature instead of the 
positive curvature  suggested by equation (8) might be realized.
A corresponding guess taken from our previous paper 
\cite{drechsler} is shown
in figure~6 (right). Anyhow, a more sophisticated study should shed light
on the "screening" mechanism of the bare Coulomb repulsion. In this context 
it is noteworthy that the arsenic-polarization (solvation) mechanism
proposed by Sawatzky {\it et al.} \cite{sawatzky,berciu} briefly discussed 
in the next 
subsection 5.1.2
seems to give a similar behaviour
as equation (8) (see figure (7) (right)).

In contrast to Haule {\it et al.} \cite{haule},
Anisimov and Shorikov {\it et al.} \cite{anisimov,shorikov} 
argue that LaOFeAs is in an
intermediate $U$ regime
but strongly affected by the value of the
Hund's rule exchange $J_{d}$.
Hence,
$\varepsilon_{\infty}$ should be strongly 
increase and the adopted relatively strong Coulomb repulsion $U_{d}=4$~eV
is not compatible with the available optical data implying
a much larger $\varepsilon_{\infty} >$ 10-15.
Thus, we are obliged to consider a weakly correlated scenario.

A somewhat similar attempt to describe the experimental reflectivity
$R(\omega)$, a reduced
plasma frequency and "bad metal" effects, i.e.\ 
 strong damping effects (first considered by Haule
\cite{haule} like in a marginal Fermi liquid )
has been undertaken in a recent study by Laad {\it et al.} \cite{laad}
using a simplified version of the dynamical mean-field theory.
Although at first glance a seemingly
similar behaviour has been obtained, quantitatively
the results are not very convincing
because (i) using $U_{\rm d}=$ 4 eV and at the same time
 ignoring the hybridization
with As 4$p$ states reduces the minimal eight-band
model to an effective five-band model
where {\it reduced} effective Coulomb interactions and in particular also
the onsite term $U$-values should be considerably reduced,  (ii)
the important screening by the background due to $pd$ and other non-$dd$ 
interband transition 
described implicitly also by
$\varepsilon_{\infty}$ in our approach has been ignored in calculating
the reflectivity
$R(\omega)$, (iii) a comparison with results from measurements on powder
samples requires a proper treatment of the anisotropic grains, for instance,
at least approximately within an effective medium theory.

\subsection{The weakly correlated As-polarization scenario}
Taking into account the criticism of strong on-site Coulomb repulsion $U_d$ 
and Hund's rule coupling
exchange $J_d$ mentioned above, the recently proposed novel, quite interesting
polarization (solvation) scenario \cite{sawatzky} for 
the pnictide components As, Te, Se, or P as seen by the Fe 3$d$ electrons
is of special interest. 
In this context we will compare our empirical values of 
$\varepsilon_{\infty}$ with the contribution of the polarizability
from As alone, $\tilde{\alpha_{\rm As}}$, as  adopted by 
Sawatzky {\it et al.} \cite{sawatzky}.
For this purpose to get a first rough estimate, 
we employ the well-known Clausius-Mossotti (also known as Lorenz-Lorentz 
relation) for the sake
of simplicity ignoring anisotropic screening effects
\begin{equation}
\alpha_{tot}=\frac{v}{b}\frac{\varepsilon_{\infty}-1}
{\varepsilon_{\infty}+2},\quad b=\frac{4\pi}{3}\qquad \mbox{or}\qquad 
\varepsilon_{\infty}=\frac{3}{1-b\alpha_{tot}/(v)}-2 > 1, 
\label{clausius-mossotti}
\end{equation}
(Note that the Lorentz factor $b=4\pi/3$ has been derived for the case of cubic symmetry.) Using for 
LaO$_{0.9}$F$_{0.1}$FeAs $\varepsilon_{\infty}$=12 and 
$v=$~140.5\AA$^3$ with two As-sites per 
unit cell we arrive at 
$\alpha_{tot}=26.35$~\AA$^3$ and $\tilde{\alpha}_{tot}=13.17$~\AA$^3$ 
per As-site already 
close to 9 to 12~\AA$^3$ adopted by 
Sawatzky {\it et al.} \cite{sawatzky}.  
For BaFe$_2$As$_2$ with $\varepsilon_{\infty}$=14.9,
$v=$~203.96\AA$^3$ with
four As sites per unit cell one obtains 
$\alpha_{tot}=$40.05~\AA$^3$ which corresponds 
to $\tilde{\alpha}_{tot}=10.01$~\AA$^3$ per As-site. Since the other atomic 
sites are expected
also to contribute to $\tilde{\alpha}_{tot}$, we briefly consider their 
influence adopting the simple expression provided by 
the polarizability additivity rule \cite{shannon}
\begin{equation}
\tilde{\alpha}_{As}=\left(\tilde{\alpha}_{tot}-\frac{1}{N_c}
\sum_i\alpha_{{\rm ion}_i}\right)/
(1-c_{\tilde{v},As}),
\end{equation}
where the sum runs over all remaining ions(atoms), $N_c$ is the number of 
As-sites per unit cell,  
and $c_{\tilde{v}, As}\sim 0.1$
denotes the atomic concentration of As-vacancies. Then
it is clear that these
ignored ionic contributions 
may somewhat change our empirical estimates of
the dominant As-contribution we are mainly interested in.
In fact, adopting for the contribution of 
Ba$^{2+}$ 1.7~\AA$^3$ and -1.0~\AA$^3$ for that of Fe$^{2+}$
\cite{dalgarno,remark} one arrives at a slightly larger 
value for 
$\tilde{\alpha}_{\rm As}\approx 10.32$~\AA$^3$,  
and a smaller value of $\tilde{\alpha}_{{\rm As}^{3-}}\approx 7.89$~\AA$^3$
is obtained
using instead the parameters 
$\alpha_{{\rm Fe}^{2+}}=1.339$~\AA$^3$ and 
$\alpha_{{\rm Ba}^{2+}}=1.57$ as recommended in reference \cite{shannon}.
Taking into account the presence of 
As-vacancies in the order of 5\% atomic percent \cite{ni}
a lower value of $\tilde{\alpha}_{{\rm As}^{3-}}\approx 8.3$~\AA$^3$ 
is obtained.

For the 
former
case of
LaO$_{0.9}$F$_{0.1}$FeAs we estimated these corrections 
due the presence of other ions, too. 
Using the following values for their polarizabilities \cite{zhaox}: 
1.13~\AA$^3$ for La$^{3+}$, 2.68  for O$^{2-}$,
3~\AA$^3$ for F$^{-}$ \cite{dalgarno}, 
and again -1~\AA$^3$ for Fe$^{2+}$ \cite{dalgarno}. As a result we 
finally arrive at an empirical value of $\tilde{\alpha}_{{\rm As}^{3-}}=
10.49$~\AA$^3$.
But using again the empirical polarizabilities recommended by 
Shannon and Fischer
\cite{shannon} with $\alpha^0_{{\rm O}^{2-}}$=1.988~\AA$^3$ and 
$\alpha^0_{F^-}$=1.295~\AA$^3$ one arrives at 
$\tilde{\alpha}_{{\rm As}^{3-}}=8.74$~\AA$^3$.
Thus, also for this different set the As-polarizabilities 
of two different materials
are large
and 
close to each other \cite{remark-arsenic}. 
Thus, the empirical numbers derived from our 
reflectivity data 
are in excellent agreement with $\alpha_{As}\approx 10$~~\AA$^3$
suggested by Sawatzky {\it et al.} \cite{sawatzky}, if the ionic 
polarizations
of all ions are taken into account.
   
\begin{figure}[t]
\begin{minipage}{8cm}
\includegraphics[width=7.9cm,angle=-90]{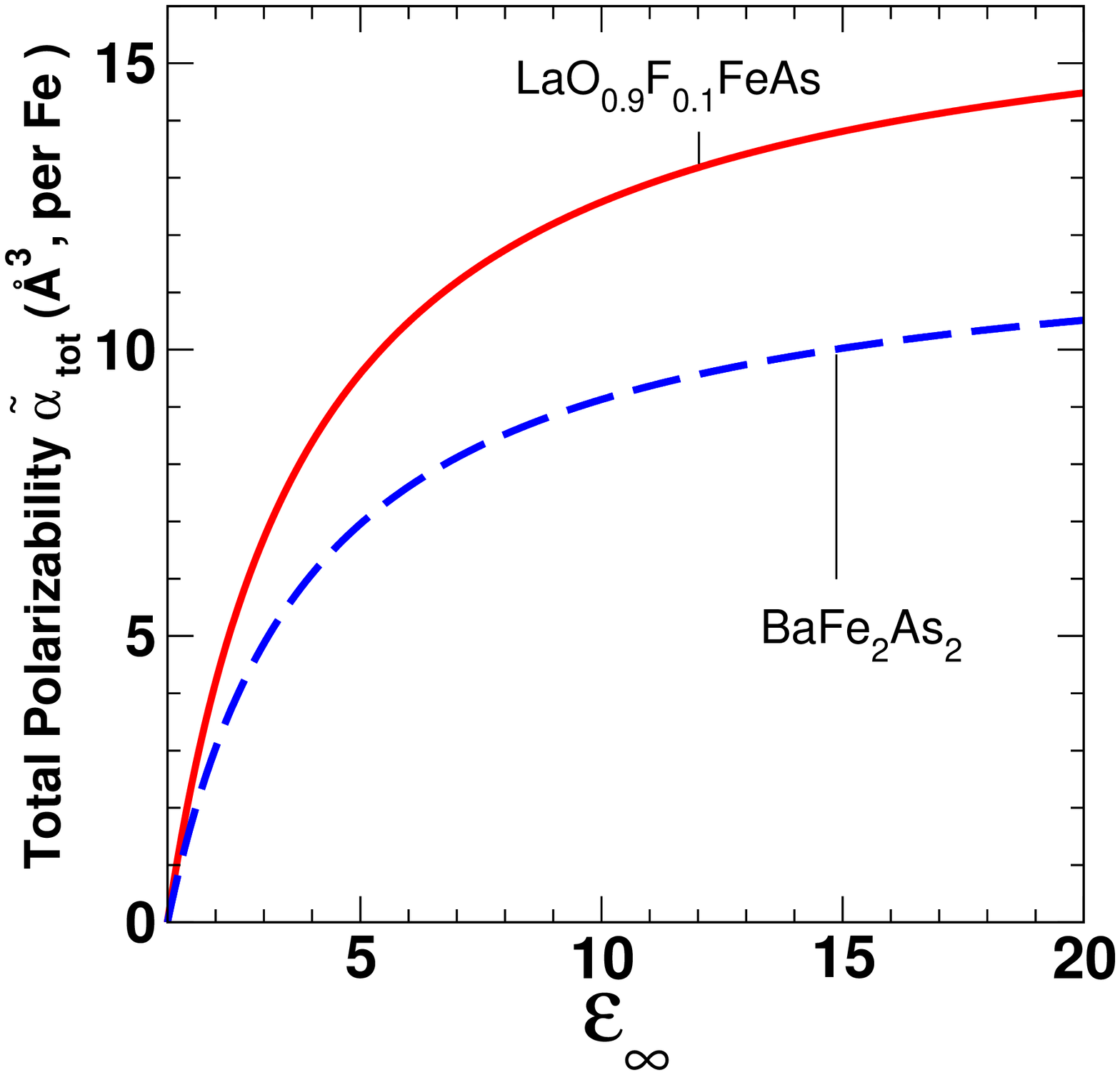}
\end{minipage}
\begin{minipage}{8cm}
\includegraphics[width=7.9cm,angle=-90]{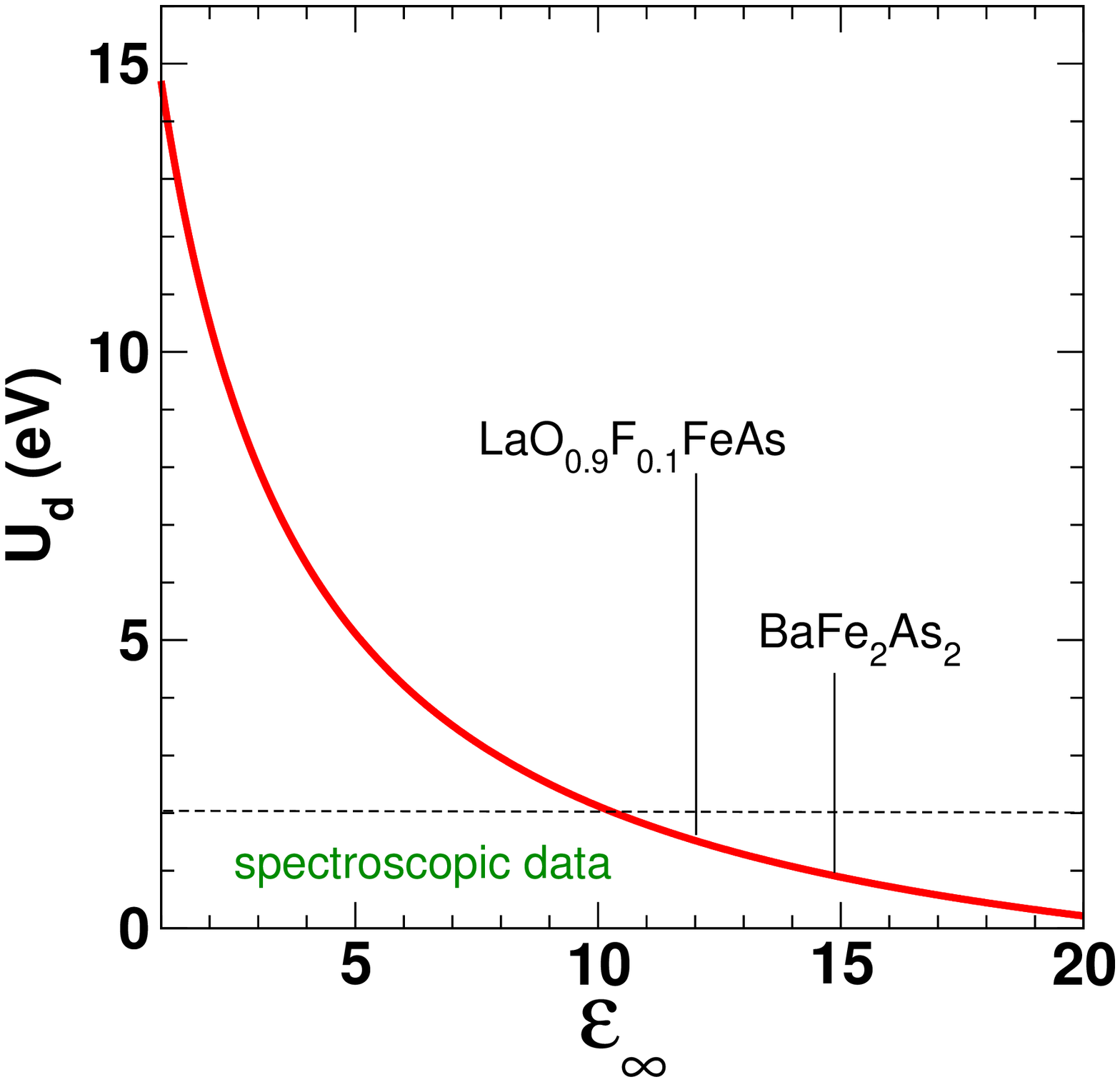}
\end{minipage}
\caption{Total polarizibilities from equation (\ref{clausius-mossotti})
(left). Schematical view of the "screened" on-site Coulomb
repulsion $U_d$ of Fe 3$d$ states 
(the so called Hubbard $U$) vs.\ background dielectric 
constant (right). For the bare case we adopted $U_{0}=14.7$~eV 
\cite{pou}. The stripe denotes the region of $U_d$ estimated
by various electon spectroscopy methods \cite{kroll,koitzsch}}
\label{Uvareps}
\end{figure}
Next, LaOFeP could be used to understand the role of polarization
effects comparing the $\varepsilon_{\infty}$ with those of the La-1111
based arsenic compounds with sufficiently higher 
superconducting and magnetic
transition temperatures,
except the parent compound itself.
Since the corresponding fit parameters
except the unscreened plasma frequency were not shown in reference
\cite{quazilbash} we are restricted to a qualitative discussion. From
the reflectivity data shown in figure 1 of reference \cite{quazilbash}   
one might ascribe the local minimum in between 4000 and 5000~cm$^{-1}$,
(i.e.\ around 0.5~eV) to the plasma edge but then one would arrive at 
a larger or close $\varepsilon_{\infty}$ for LaOFeP compared with 
the arsenic counterpiece. Therefore we suggest that the actual plasma edge 
is masked
by an interband transition or by a P-vacancy induced absorption like from the 
As-vacancies
in two of the above considered arsenide
systems. In fact, for a sequence of 
polarizarion differences
$\tilde{\alpha}_{\rm As}-\tilde{\alpha}_{\rm P}=0,1,2,3$~\AA$^3$ we estimate 
(predict)
from equation (9)
$\varepsilon_{\infty,{\rm P}}=14.4$, 10.2, 7.7, 6.1, respectively. Then 
the plasma edge would be near 0.487~eV, 0.58~eV, 0.67~eV, 0.75~eV, where
$\varepsilon_{\infty,{As}}=11.5$ and $v_{\rm P}=133.75$\AA$^3$ have been used.
In addition we have assumed that the polarizabilities of the other ions 
La$^{3-}$, O$^{2-}$, and Fe$^{2+}$ remain
unchanged, i.e.\ they cancel out in the difference 
$\tilde{\alpha}_{\rm As}-\tilde{\alpha}_{\rm P}$. 
 A P-vacancy scenario might also explain the 
suppression of an antiferromagnetically ordered state in favour of 
superconductivity as observed
 in LaOFeP below 6K. If indeed $\varepsilon_{\infty,{\rm P}}$ is sufficiently
smaller than 11 to 15 found here for two arsenide compounds, we would
predict also a somewhat larger $U_d$ value for LAOFeP. Both predictions can 
be checked 
experimentally.  

To get some insight in the accuracy of equation (\ref{clausius-mossotti})
and to check the polarizability of the next important most competing
ionic component O 
we considered the cuprates La$_2$CuO$_4$ and Li$_2$CuO$_2$ in the same way 
using in addition 
0.2~\AA$^3$ for the polarizability of Cu$^{2+}$ and 0.0286~\AA$^3$ for Li$^+$
summing up all ionic polarizabilities one obtains 
$\alpha_{tot}=24.28$~\AA$^3$ and 11.31~\AA$^3$, respectively,
(both with two formula units per unit cell). As a result 
with unit cell volumes of 
$v=187.89$~\AA$^3$ and $v=98.42$~\AA$^3$
realistic $\varepsilon_{\infty}$-values of 4.54 and 3.74, 
respectively,
are obtained from equation (\ref{clausius-mossotti}).
Using the set of polarizabilities proposed by Shannon
and Fischer \cite{shannon} $\alpha^0_{O^{2-}}=1.988$~\AA$^3$,
and $\alpha_{Cu^{2+}}=1.23$~\AA$^3$, 
we arrive at 
$\alpha^0_{tot}=10.524$~\AA$^3$ and $\varepsilon^0_{\infty}=3.434$ 
in the free ionic approximation \cite{remarkshannon}.
 The analysis of the loss function -Im$\left[1/\varepsilon(\omega)\right]$ for
Li$_2$CuO$_2$ within the framework of a five-band Cu3$d$-O$2p$ 
extended Hubbard model
\cite{malek} reveals with $\varepsilon_\infty =3.6$ a very close value.
The calculated ionic charges $Q_{Cu}=1.797e$ and $Q_O=-1.972e$ are also
very close to the idealized ionic picture with values of 
2$e$ and -2$e$,
respectively. 
 
In the cuprates 
the background dielectric constant $\varepsilon_{\infty}$ is
 dominated by the polarizability of $O^{2-}$, i.e.\ by the 2$p$-component
similarly as in our case by the As 4$p$ states.
Anyhow, in spite of the surprisingly good description in both cases 
more sophisticated
calculations beyond the isotropic cubic approximation 
used in deriving
equation (\ref{clausius-mossotti}) are nevertheless
highly desirable \cite{yatsenko,dolgov} in view of the importance
of the polarization problem for the elucidation of the 
mechanism of superconductivity in
the iron pnictides.
In this context we would like to note that our estimated empirical
values of $\alpha_{As} < 12$\AA$^3$ are probably not large enough to 
ensure a bipolaronic formation at low temperature, i.e.\ a 
Bose condensation-like scenario of 
superconductivity in the iron based pnictides such
as proposed by Sawatzky {\it et al.} \cite{sawatzky,berciu}
This  objection is also supported by the observation of a nearly
linear temperature dependence of the upper critical field
near $T_c$  instead of an expected 
pronounced positive curvature generic for bipolarons $B_{c2}(T) 
\propto (T_c-T)^{1.5}$ \cite{alexandrov}, 
the Pauli-limiting behaviour 
observed in some cases \cite{fuchs1,fuchs2} 
as well as the clear observation of Fermi surface related features
such as the de~Haas~van~Alphen (dHvA) effect in LaOFeP \cite{coldea} and last 
but not least the
numerous angle resolved photoemission experiments (ARPES) \cite{ding,borisenko1}.
Anyhow, our conclusions about unstable bipolarons 
should be taken with some cautions in view of the 
approximate 
character of equation (\ref{clausius-mossotti} and the fact that our 
estimated 
polarizabilities are determined from optical frequencies around the 
screened plasma 
frequencies whereas the dielectric polarizibilities entering the 
original Clausius-Mossotti
equation are determined usually in the range of 1kHz-10 MHz and include both
 ionic and electronic components. The extrapolation of our results to that
quasi-static limit even below typical phonon-frequencies remains an open 
and interesting
question worth
to be considered in future. We admit that such ionic effects (or lattice 
effects in general) might support the electronic polarization effects and 
favour finally the formation of bipolarons, too. However, even when bipolarons
are stabilized it is unclear what will their mass ? Too heavy bipolarons
is a general serious problem for most bipolaronic scenarios to explain rather 
high-$T_c$ values. Another possibility would be also the so-called boson-fermion
scenario (or $s$-channel scenario) where real bipolarons occur not as prepaired 
pairs already above $T_c$ but occur only below $T_c$ where they coexist with 
Cooper pairs (see e.g.\ reference \cite{mamedov} and references therein). 

In order to illustrate the polarization effect on the Coulomb repulsion on Fe
sites 
we consider again the relationship between $U_d$ and $\varepsilon_{\infty}$
employing equation (\ref{clausius-mossotti}). According to 
reference \cite{sawatzky}
the actual on-site Coulomb repulsion $U_d$ in a multiband Hubbard-model 
(for the sake of simplicity 
we ignore slight differences between the various involved five Fe 3$d$-orbitals)
is given by the difference 
\begin{equation}
U_d=U_0-2E_p, \qquad E_p\sim 0.5\sum_iE_i^2,
\end{equation}
where the polarization energy $E_p$ reflects the electric field 
produced at all surrounding polarizable sites due to the additional charge
related to the on-site double occupancy and $U_0$ is the bare Coulomb repulsion
for which we adopt 14.7~eV \cite{pou}. In the point charge approximation
one has $E_p=\sum\tilde{\alpha}_{ji}e^2/R^4_{ij}$ resulting in 
$2\tilde{\alpha} e^2/R^4_{Fe,As}=0.868\tilde{\alpha}$/[\AA$^3$]~eV
in the nearest neighbour (NN) approximation ignoring the influence of other ions, 
where $R_{Fe,As}\approx 2.4$~\AA \ denotes the NN arsenic-iron distance.
Instead of taking into account also the dipole-dipole interaction 
and summing up all contributions
within a single FeAs layer as it was done in 
reference \cite{sawatzky} in the point charge approximation, we simply assume
that the polarization energy $E_p$ scales linearly with $\alpha$ where 
the constant of 
proportionality $u_1$ has been 
determined from the condition to reproduce approximately
our DMFT based result of $\varepsilon_{\infty}=5.4$
for $U=4$~eV given in the previous subsection 4.1. 
Using equations (9) and (11) the resulting 
phenomenological curve shown
in figure 7 (right) reads
\begin{equation}
U_d=U_0-u_1\frac{3\tilde{v}}{4\pi}\frac{(\varepsilon_{\infty} -1)}{(\varepsilon_{\infty} +2)},
\end{equation}
where $\tilde{v}$ denotes the volume per iron site. This way we 
arrive at $U_d \approx 1.5$~eV
for LaO$_{0.9}$F$_{0.1}$FeAs and $U \approx $ 1~eV for BaFe$_2$As$_2$ using the 
same DMFT based coefficient $u_1$ also for the latter compound. 
Thus, our estimated $U_d$ values are in both cases
{\it smaller} than  the total band width  
$W \sim $2~eV.
Taking into account the quasi-degeneracy of all five Fe 3$d$ orbitals involved 
one comes to the conclusion
that repulsive correlation effects should play a minor role except maybe for the 
magnetic exchange interaction affected by the superexchange and the Hund'rule 
couplings as well.
The same reduced size of the on-site Coulomb repulsion
has been estimated based on various electron spectroscopy measurements
\cite{kroll,koitzsch}.
Note that residual Coulomb interactions of $U_{eff} =320$ to 380~meV 
and a Hund's rule coupling
$J_{eff}=70$~meV have been used for instance by Korshunov 
and Eremin
\cite{korshunov,klauss}
within a four-band theory for the description of thermodynamic properties
related to the nesting derived spin density wave (antiferromagnetic) ground state
and its
magnetic excitations of the undoped parent compound LAOFeAs and other 
the iron pnictides 
(We remind the reader that the extended Hubbard model including
 besides all five Fe 3$d$
also the As 4$p$ orbitals is a 16-band model.) Without doubt our 
estimated strongly
reduced $U_d\sim $1~eV-values might be helpful to understand microscopically
the low value of the former, especially, if a reduced, but yet sizable, 
value of the NN intersite Fe-Fe Coulomb
interaction $V$ is taken into account: $U_{eff}\sim U-V$, provided 
a repulsive $V>0$ is still obeyed in contrast to an expected {\it attractive} 
$V<0$ adopted within the bipolaronic scenario of 
Sawatzky {\it et al.}, where the polarization 
$\tilde{\alpha}_{\rm As}\geq 12$\AA$^3$ is so strong to change its sign, 
i.e.\ $V$ would be over-screened. In addition, $U_{eff}$
is also reduced in the process of mapping down the 16-band model "integrating out" the
As-4$p$ states \cite{anisimov} exhibiting a moderate hybridization  with 
the Fe 3$d$ states. Anyhow, the extremely small value 
of $J_{eff}$ compared with expected
0.7 to 1 eV remains as a puzzle to be resolved. 
(Since 2$J$ is given as the difference
of the both onsite (intra-atomic) $U$ and the 
intra-atomic-interorbital $U'$ and to first approximation
both are expected to be reduced in the same way by the polarization of 
the surrounding ions.). Possibly, the explicit account of an additional 
direct ferromagnetic 
exchange interaction $K_{pd}$ might be helpful to
resolve this puzzle.
$K_{pd}\sim 200$~meV  was introduced for the two dimensional corner-shared
cuprates by Hybertsen {\it et al.} \cite{hybertsen}.  
For instance, in almost all 
edge-shared cuprates such as the above mentioned Li$_2$CuO$_2$ or in 
Li$_2$ZrCuO$_4$ \cite{drechsler07}
with Cu-O-Cu bond angles near 90$^\circ$ (just like as the 
Fe-As-Fe bond angle in the iron pnictides considered here) $K_{pd}$ 
amounts about 50 to 80~meV \cite{malek,drechsler07}. There $K_{pd}$
is decisive for the observed ferromagneic NN-exchange interaction
being much more importantant than the ferromagnetic coupling provided 
by the Hund's rule coupling $J_p$ on the intermediate oxygen O 2$p$ orbitals.
The reason for that is that $K_{pd}$ affects the total exchange already in 
the second order of the perturbational  theory whereas the Hunds' rule $J_p$
coupling appears
only in fourth order (four small $pd$ transfer intregrals) \cite{eskes}.   

Also a negative $U$-scenario as proposed in reference \cite{nakamura}
to solve the puzzle of the too large calculated magnetic moment
in the framework of LDA+$U$ seems to be
unlikely. According to our estimate 
$\varepsilon_{\infty} > 25$ would be required 
(see figure \ref{Uvareps}). Anyhow, a detailed 
comparison with other metalic compounds which exhibit
large $\varepsilon$-values would be interesting also in view of
searching for new superconducting compounds and for systems with other 
exotic ground states.

First estimates of the corresponding polaronic effect
result in mass renormalization by a factor of 2 in accord 
with our findings for $\Omega_p$
and those from ARPES data for 122 superconductors \cite{ding} 
and de Haas van Alphen studies 
for LaOFeP \cite{coldea}.
Thus, the mentioned above strongly reduced Coulomb repulsion $U_d$ can be 
understood based on empirical considerations for the ionic polarizabilities and 
is in accord with empirical findings analyzing various
electron spectroscopy data \cite{kroll,koitzsch,anisimov}.

\section{\bf The {\it el-boson} coupling strength and constraints for
pairing mechanisms}
Naturally, the most interesting problem of the pnictide superconductivity
is the elucidation of its mechanism. At the present relatively poor
knowledge
and still
not very high sample quality (with respect to impurities and
other defects)
reflected for instance by the high residual resistivities 
$\rho_0$
and an unknown
mesoscopic structure due to a pronounced tendency to an
electronically driven phase
separation \cite{aczel,keimer} at least for the (122)-family,
a comprehensive
theoretical description is impossible. In such a situation one may adopt
various simplified scenarios (with increasing complexity) and look for
qualitative checks like bipolaron condensation vs.\ Cooper pair formation, 
the symmetry of the order parameter(s), the
coupling strength and the effect of disorder in order to discriminate
at least a part of the 
huge amount of already proposed scenarios.

\subsection{One-band Eliashberg analysis}
We start with the simplest possible version of an effective single-band
theory where the formation of a condensate of Cooper pairs is assumed.
(The case of prepaired bipolarons can be excluded by the nearly
linear $T$-dependence of the upper critical field near $T_c$
(see Fuchs {\it et al.} \cite{fuchs1,fuchs2}.)
Here the  mass of the paired charge carriers is  affected by
the virtual retarded interaction with collective excitations (bosons) 
being present in any solid. The corresponding 
{\it el-el}  interaction may be attractive or repulsive, 
supporting
or weakening a specific type (symmetry) of pairing in addition to 
the unretarded
Coulomb interaction which may be also repulsive or attractive under special 
circumstances \cite{kulic}.  
Thus, from the standard 
Eliashberg-theory applied to the case of type-II superconductor with
weak or intermediately strong
coupling ($\lambda \leq 2$)\cite{waelte} one
obtains for the penetration depth $\lambda_L(0)$ 
at zero temperature using convenient units
\begin{equation}
\sqrt{\varepsilon_{\infty}} \omega_p \mbox{[eV]}\frac{\lambda_L(0)\mbox{[nm]}}{\mbox{197.3nm}}=\sqrt{(n/n_s(0))(1+\lambda_{tot}(0))(1+\delta)},
\label{singlepen}
\end{equation}
where $\lambda_L(0)$[nm] is the  experimental
$(a,b)$-plane penetration
depth extrapolated to $T=0$.
$\Omega_p=\sqrt{\varepsilon_{\infty}} \omega_p $ denotes
the empirical unscreened plasma
frequency shown in table 2, $n_s(0)$ is the density
of
electrons in the condensate at $T=0$
and $n$ denotes the total electron density of all conduction bands
which contribute to the unscreened $\Omega_{p}$,
and $\delta$ is the disorder parameter
\begin{equation}
 \delta=0.7\frac{\gamma_{imp}}{2\Delta(0)}
\equiv 0.7\frac{\gamma_{imp}}{RT_c}\approx \frac{\gamma_{imp}}{5T_c},
\label{delta}
\end{equation} 
which is small but finite even in the quasi-clean limit. 
$R$ denotes the gap-to $T_c$ ratio
$R=2\Delta(0)/T_c$ which in moderately strong coupled superconductors 
slightly exceeds the BCS-value $R_{BCS}=3.52$.  
Notice that all three factors, the strong coupling correction described 
by the 
mass enhancement factor $Z=1+\lambda$,
the disorder $D=1+\delta$, as well as the condensate occupation parameter
$N=n/n_s$  under the  root obey the inequality $\geq$~1.
Let us note that the possibility to extract the {\it el-boson}-coupling 
constant
from a comparison of the high frequency optical response 
with the renormalized 
low-energy plasma frequency which determines the superfluid density
has been proposed (in the clean-limit $D\equiv 1$ and for the BCS case $N=1$) 
already in 1987 by Gurvitch and Fiory \cite{hirsch-pen} and again in  
1992 by Hirsch and Marsiglio \cite{hirsch}. But to the best of our 
knowledge there were no 
systematic applications to real superconductors. 
A first step in this direction is presented in subsections 5.1.1 and 
5.1.2. Note that  equation (\ref{singlepen}) can be generalized taking into 
account
two unusual situations which, however, may be present for certain pnictides
as discussed below in subsection 5.1.2 in the context of 
LaOFeP \cite{quazilbash} in more detail: 
(i) some residual mass renormalization is still present  at room temperature
in the frequency region of the screened plasma energy $\sim$ 0.4 to 0.5 eV 
i.e.\ $\lambda(\omega=\omega_p, T=300$ K) is not very small. This may happen
when the spectral density of the coupled boson extents to very high-frequencies
and (ii) the superconductivity is somewhat suppressed due to fluctuations
of a competing neigbouring (e.g.\ magnetically ordered) phase. To take these 
effects
into account we may write:
\begin{equation}
Z=\frac{1+\lambda_{tot}(0)}{1+\lambda_{tot}(\omega_p,T)}, \ 
\mbox{or}\ \lambda_{tot}(0)=\frac{Z-1}{1-r}, \ \mbox{where} \quad r=\frac{\lambda_{tot}(\omega_p,T)}{\lambda_{tot}(0)},
\label{generalz}
\end{equation} 
and
\begin{equation}
D=(1+\delta)(1+f),
\label{deltafluct}
\end{equation}
where $f\geq 0$ is a phenomenological parameter which measures the 
fluctuational influence 
of all competing phases. The latter effect occurs beyond the standard 
mean-field approximation
and a for a treatment of all instabilities (various superconducting phases
with different symmetry of the order parameter ($s,s^{\pm},p,d$-waves), 
SDW's, ferromagnetism, and CDW's, etc.)
on equal footing. Since such a corresponding comprehensive microscopic
 theory is not avalaible at present,  
$f$ has to be treated as a phenomenological parameter.  

In classical BCS-type superconductors at $T=0$ all quasiparticles 
participate in the superfluid condensate, i.e.\ $N=1$. In complex 
multicomponent 
superconductors
like underdoped cuprates the possibility of $n_s < n$ has been discussed 
by several authors (see e.g.\ R\"ufenacht {\it et al.} \cite{ruefenacht}  ).
In the present case it might also happen in the case
of coexisting CDW or SDW orderung with superconductivity
as observed for instance in transition metal dichalcogenides such as 
NbSe$_2$ or in clean magnetic borocarbides (HoNi2B2C). 
The well-known classical superconductor Nb which much stimulated the
development of the strong coupling theory
provides an almost ideal
possibility to illustrate the power of equation (\ref{singlepen}). 
Adopting $N=1$ we may examine the case of very clean samples with 
resistivity 
ratios RRR up to $\sim 2\cdot 10^4$ (!)\cite{berthel} for which a small 
disorder parameter $\delta$ is expected. In fact, with $\lambda_L(0)=31.5$~nm
\cite{auer}, 32~nm \cite{anlage},
34.3~nm \cite{berthel}, 35~nm \cite{karakozov} and the calculated LDA-value of
$\Omega_{p}=$9.24~eV \cite{romaniello} or the empirical value
of 9.2~eV~\cite{karakozov} 
and $\lambda=1.12$, $T_c=9.22$~K \cite{zheng} 
one arrives at $\delta=0.026$, 0.059 and 0.217, respectively,
which point to a very clean-limit regime. In fact, 
using the experimental gap to $T_c$ ratios of 4.1 \cite{anlage,pronin} 
and 3.79 \cite{berthel} these $\delta$-values correspond very small 
scattering rates 
of 1.4~K, 3.2~K, and 10.8~K, respectively, only.
Based on these results we will denote a regime with $\delta <1$
as a quasi-clean limit.

Turning back to LaO$_{0.9}$F$_{0.1}$FeAs we adopt also the BCS relation 
$n$=$n_s$
and $\delta \approx $0.93 being still in
 the quasi-clean limit at $T=0$. Then, using
$\varepsilon_{\infty}$~=~12 \cite{boris} and $\lambda_L(0)$~=~254~nm
\cite{luetkens} or 242~nm \cite{drew} one estimates
$\lambda_{tot} \sim 0.6$ and 0.45, respectively, i.e.\
the superconductivity is in a weak coupling regime
for our samples with a bulk
$T_c$~=~26~K (according to the $mu$SR-data \cite{luetkens}
while the resistivity yields a slightly larger value of 27-28 K).

\begin{figure}[b]
\hspace{2cm}
\includegraphics[width=9.0cm]{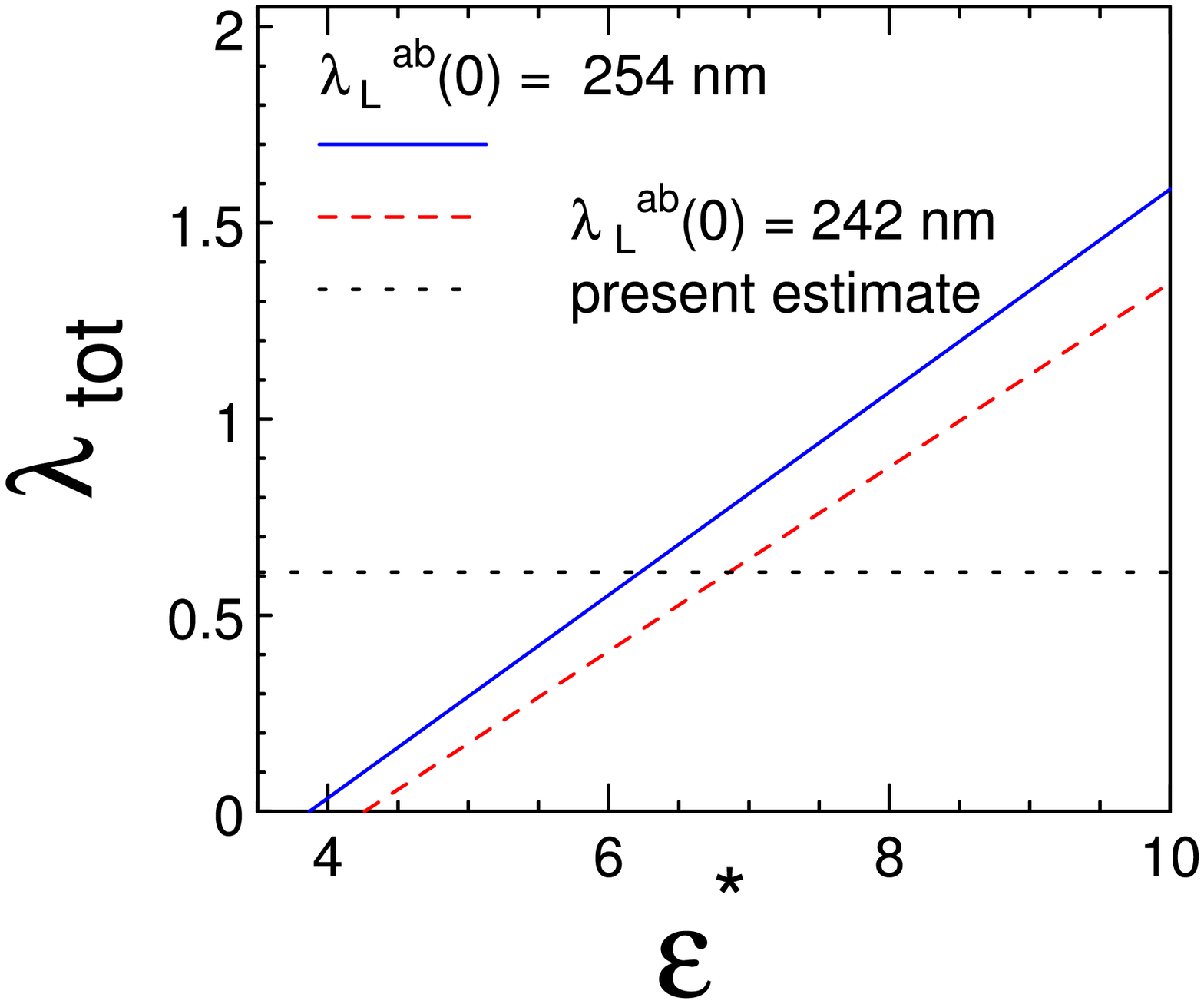}
\caption{Empirical relation between
the total coupling constant from the mass enhancement
entering the penetration depth $\lambda_L(0)$ vs 
polarizability related renormalized
dielectric background constant $\varepsilon^*$.
The in-plane penetration depth values $\lambda^{ab}_L(0)$ 
have been taken from references
\cite{luetkens,drew}.  
}
\label{f8}
\end{figure}

The effect of the dielectric background constant 
renormalized by both the disorder and the actual condensate density  
$\varepsilon^*=\varepsilon_{\infty}/DN$ on the coupling strength
is shown in figure~\ref{f8}.
Note that a substantial
impurity scattering absent in the quasi-clean limit would
further
reduce $\lambda_{tot}$.
We note once more that our empirical $\Omega_p \approx$~1.36~eV
  differs 
from the LDA-prediction of 2.1~eV for the
nonmagnetic ground state  within the LSDA.

\subsubsection{Comparison with other pnictide superconductors}
In the present context it is interesting to apply our analysis also to 
other iron based pnictide superconductors 
for which the penetration depth is known.
Khasanov {\it et al.} \cite{khasanov-Se} reported $\mu$SR measurements 
for the related FeSe$_{0.85}$ system with a relative low $T_c=8.26$~K
(probably related to the smaller polarizability of Se compared with As). 
They found from their $\mu$SR-data a penetration depth
  $\lambda_L(0)= 406$~nm. Adopting a typical plasma frequency of about 
1.3 to 1.5~eV (compare table 2),  one arrives at $ZDN \approx 7.16$~to~9.53.
For a weak to intermediately strong {\it el-boson} coupling strength
$\lambda_{tot}=0.5$~to~1 one is left with a relatively large factor
$DN \geq$ 3.58 to 4.76 
i.e.\ either the system under consideration is strongly disordered
or affected also by fluctuations of a 
competing phase (see equation (\ref{deltafluct})).
Alternatively this factor is enhanced due-to $N>1$, i.e.\ 
the condensate 
density $n_s$ is much smaller than the total electron density $n$.
Finally, Li {\it et al.} \cite{li} studying Ba$_{0.6}$K$_{0.4}$Fe$_2$As$_2$
($T_c=37$~K) reported $\lambda_L(0)=200\pm 8$ nm adopting a dirty limit 
scenario  in order to describe the observation of a gap in their 
reflectivity measurements. Then, in fact, adopting a relatively large 
value of 
$\Omega_p \approx 1.8$ to 2.1~eV such as 
suggested in reference \cite{yang} for the closely related system
Ba$_{0.55}$K$_{0.45}$Fe$_2$As$2$ one estimates from equation (\ref{singlepen})
 $ZDN$= 3.28 to 4.47 and there is room for some disorder
for weak to intermediately strong coupling $Z\sim 1.6 $ to 2 provided $N=1$.

The most interesting from the disorder and high-field properties  point of 
view is the closely related system given by
the arsenic-deficient compound LaO$_{0.9}$F$_{0.1}$FeAs$_{1-\delta}$
with improved superconducting properties near $T_c\approx 28$ K 
\cite{fuchs1,fuchs2}.
The analysis of
the Pauli limiting (PL)
behavior for the closely related LaO$_{0.9}$F$_{0.1}$FeAs$_{1-\delta}$
system yields a similar value
$\lambda \approx 0.6$ to 0.7 as derived from the 
Eliahberg-theory  corrections to the BCS (called often 
``strong-coupling'' corrections even for small or moderate values of
$\lambda$ )
for its
PL field $B_P$(0)~=~102~T \cite{fuchs1,fuchs2}.
Quite interestingly, the observed kink in the electron dispersion in the
ARPES data by Wray {\it et al.} \cite{wray} for the 122 
(Sr,Ba)$_{1-x}$(K,Na)$_x$Fe$_2$As$_2$ can be described with an effective
$\lambda \stackrel{>}{\sim} 0.6$ of the same order as that found for 
the La-1111-system \cite{drechsler} and adopted
above for the corresponding rare earth system (see figure 9) and discussed 
in more detail there.

In spite of relatively large error bars it is noteworthy that an 
Uemura-plot type relation between the superconducting transition 
temperature and the squared of the empirical {\it unscreened} plasma energy 
$\Omega_p$
seems to hold (see figure 9)
although the 1111-pnictides occur on {\it another}   
curve than the 122-pnictide superconductors \cite{hiraishi}.
The 1111-pnictides are more or less 
close to the hole-doped cuprates on an almost
linear curve but with an additional negativ constant.
At present  it is unclear what does this penomenological 
result imply? Is there a critical carrier concentration 
necessary for the occurence of superconductivity at variance
with the Uemura-plot prediction? 
Or does it mean that for the present crystal structure types
(1111, 122, and 111)
very low carrier phases are simply not realized for chemical 
reasons and the vanishing $T_c$ near $\Omega_p \sim 1$~eV
is caused by other reasons {\it not}  directly related to the carrier 
density? In the La-1111-family the smallest $\Omega_p$ occurs just
near the phase boundary to the SDW-phase.
 
The 122-systems can be described by another linear relation but 
also with a negative constant. 
Note that the stoichiometric compounds 
LAOFeP \cite{quazilbash} and NaFeAs\cite{parker}
 occur
 far from these two curves for the 1111 and 122 As-based families
Further experimental and 
theoretical work is required to settle these points. A solution might 
be also helpful to understand better the mechanism of superconductivity
in iron based pnictide superconductors. In the context of a (possibly)
somewhat smaller $\varepsilon_{\infty}$ and also a larger $U_d$ for LaOFeP,
its lower $T_c=6$~K as compared with values of
18~K for
LiFeAs or 26~K for LaO$_{1-x}$F$_x$FeAs might be a relevant hint
for the elucidation of the 
still unknown mechanism of superconductivity in Fe pnictides, too. 

The measured in-plane penetration depths of two LiFeAs powder samples 
with superconducting transition temperatures $T_c=$ 16 and 12~K, respectively,
amount
$\lambda){L}(0)=195$~nm and 244~nm \cite{pratt}. Since to the best of our 
knowledge optical data are still not available we estimate $\Omega_p=$1.77~eV
using equation (7) and the reported lattice constant $a$= 3.77 \AA . The 
isostructural NaFeAs exhibits a larger lattice constant $a$=3.947 
and a lower $T_c$=9~K \cite{parker}.

LAOFeP deserves special attention due to its unusual properties
and the relatively large amount of experimental single crystal data 
available for a phenomenological comprehensive
analysis \cite{quazilbash,fletcher,analytis}. 
The penetration depth exhibts a quasi-linear temperature
dependence pointing to unconventional superconductivity with nodes
in the order parameter \cite{fletcher} which points possibly to a different
superconducting phase compared with the nodeless FeAs based 
superconductors. From the measured in-plane penetration depth
$\lambda_L(0)=240$~to~250~nm \cite{fletcher} and $\Omega_p=1.85$~eV
we estimate a large value for $ZDN \approx 5.06$. Ignoring the possibility
of $N>1$ and adopting $\lambda_{tot}\leq 1$ as suggested by the 
dHvA-data \cite{coldea} and ARPES data \cite{lu}, one is left 
with $D\sim 2.5$ 
to 3.4. Since the
measurements were performed on very clean single crystals \cite{fletcher}
a large disorder parameter alone seems to be not very likely to explain
the data. Instead we assume  a sizable contribution of {\it fluctuations}
measured by our phenomenolgical parameter $f$ 
(see equation (\ref{deltafluct})) which is thought to be caused by the 
presence of at least one
competing superconducting and/or magnetic phase(s). In this context 
we note that the possibility of enhanced penetration depths due to the 
vicinity of 
fluctuating magnetic phases have been assumed very recently
also in reference \cite{vorontsov}. Finally, we note that the 
unsaturated temperature
dependence of the optical mass renormalization 
shown in the inset of figure 3 (panel b)
in reference \cite{quazilbash} points to a residual 
$\lambda(\omega=\omega_p,T=300{\rm K})$ introduced in equation 
(\ref{generalz})
which might be caused by a very broad
spectral density (Eliashberg function) $\alpha(\omega)F(\omega)$ 
with non-negligible high energy contributions. Adopting $r(300{\rm K})$=
0.25 to 0.33 would be helpful to reduce the differences of mass renormalization
as seen in optical data ($\lambda \approx 0.5$) compared with dHvA \cite{coldea} 
and ARPES data \cite{lu} pointing to $\lambda \approx 1$. The account of
$r$ in $Z$ would further increase $D$ and the phenomenological 
fluctuation parameter $f$. If such an effect would be present also in other
pnictide superconductors, our estimated coupling constants would be enhanced to 
$\lambda_{tot}\sim 1$.

 Anyhow, whether the quantity $\Omega_p$ is related 
in fact to the
superfluid density $n_s$ remains unclear. Due to some similarity 
with the original Uemura-plot
one might conclude that all free charge 
carriers are in the condensate and the renormalization by the {\it 
el-boson} interaction should be rather similar. 
Then the only free parameter within a single band approach is the 
disorder parameter $\delta$.
In this approach the rare earth 1111 compounds with 
the highest $T_c$-values $ \sim 40$ to 56~K
exhibit also the weakest disorder.
For instance, adopting the same $\lambda_{tot}=0.61$ as for
LaO$_{0.9}$F$_{0.1}$FeAs discussed above we take into account the increase
of $\Omega_p$ due to the shortening of $a$ as modelled by equation 
(\ref{plasma-scaling}) and
arrive at $\delta \approx$~0.5. 
This is a clear indication for the presence of multiband effects.
Finally, disorder 
affects mainly 
that band with the slow particles to which the penetration depth $\lambda_L$
is not very sensitive (see below).
\begin{figure}[t]
\hspace{2cm}
\includegraphics[width=12.cm]{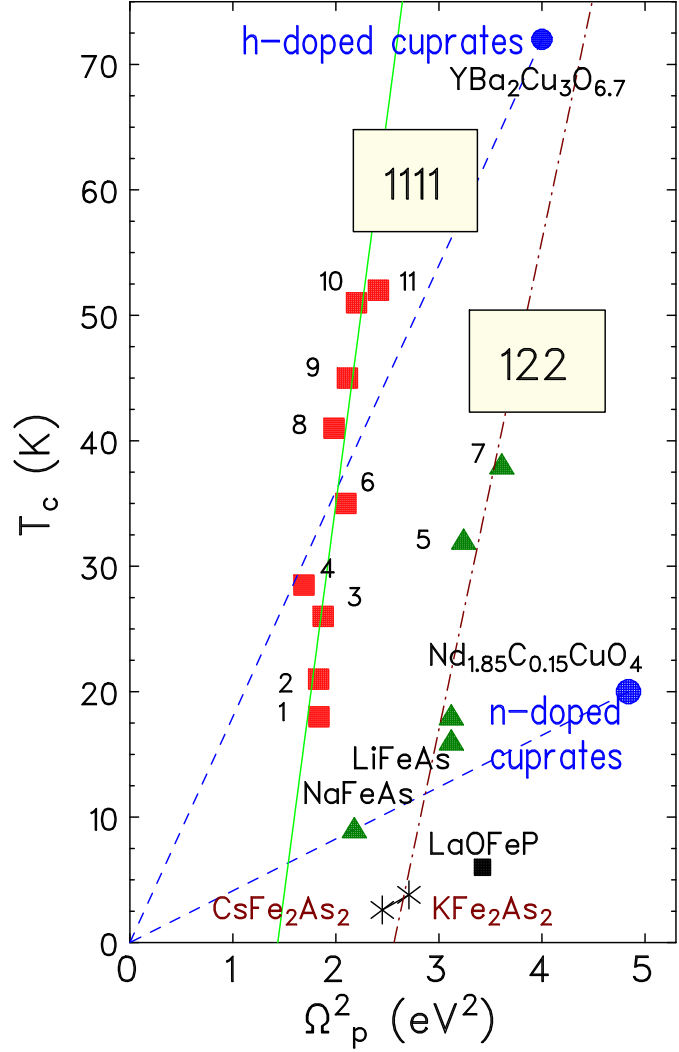}
\caption{Uemura-plot type relation between the superconducting 
transition temperature $T_c$
and the empirical squared unscreened plasma frequency 
{\it not renormalized} by the {\it el-boson} interaction 
as derived from 
optical and $\mu$SR measurements for selected Fe-pnictide superconductors.
The shown materials are LaO$_{1-x}$F$_{x}$FeAs (1- $x$=0.06 \cite{hiraishi}, 
2-$x=0.075$ \cite{luetkens} 3- $x$=0.1 \cite{drechsler})
4- LaO$_{0.9}$F$_{0.1}$FeAs$_{0.9}$ (present work),  
5- Ba$_{0.55}$K$_{0.45}$Fe$_2$As$_2$ \cite{yang},
6 - Ba$_{0.6}$K$_{0.4}$Fe$_2$As$_2$ \cite{hiraishi},
7 - CeO$_{0.84}$F$F_{0.16}$FeAs \cite{carlo}, 
8- NdO$_{0.88}$F$_{0.12}$FeAs \cite{carlo}
9 - Sm$_{0.82}$F$_{0.18}$FeAs  \cite{drew}, 
10- NdO$_{0.85}$FeAs \cite{khasanov}
11 - SmO$_{0.85}$FeAs \cite{khasanov}. The points 7-11 and LiFeAs are predictions
extrapolated from equation (\ref{plasma-scaling})
 using the actual lattice constant $a$.
}
\label{tcplasma}
\end{figure}
Finally, we note that in the related nickel arsenide 
LaO$_{0.9}$F$_{0.1}$NiAs a 
coupling constant of $\lambda \approx 0.93$ has been found from a
single-band  
Eliashberg analysis of its thermodynamic properties \cite{Li}.

\subsubsection{Comparison with other exotic superconductors
\label{comparison}}
Finally, it is interesting to compare our results for the 
Fermi surface averaged total coupling constant $\lambda_{tot}$ of the 
Fe-pnictides given above 
with a similar analysis for other more or less exotic 
superconductors where the symmetry of the order parameter may be 
unconventional and in addition to phonons also other bosons such as 
crystal field excitations and/or spin fluctuations
may play a crucial role. 

Let us start with extended $s$-wave
superconductors with intermediately strong {\it el-ph} coupling
$\lambda_{el-ph}\sim 1$ for which the necessary experimental data 
is available
(a corresponding comparison with LDA-based bandstructure calculation
will be given elsewhere). We start with

(i)
the nonmagnetic borocarbide YNi$_2$B$_2$C ($T_c=15.6$~K). 
According to Widder {\it et al.}
\cite{widder-YNi2B2C} one has $\tilde{\Omega_{p}}=4.25$ eV 
for the plasma frequency already renormalized by the {\it el-ph}
interaction and $\lambda_{opt}=1.2$
in accord with the Fermi surface averaged 
$\lambda=1.05$ from specific heat data \cite{drechsler-YNi2B2C}. 
Using the 
penetration depth 
$\lambda_L(0)\approx \lambda_L(T=3\mbox{K})=$ 58~nm from 
the $\mu$SR data by Ohishi {\it et al.} 
\cite{ohishi-YNi2B2C}, one arrives at $(1+\delta)=1.56$. Hence, 
a quasi-clean limit value such as 
$\delta \approx 0.7$ estimated from the low-temperature 
optical data
of Bommeli {\it et al.} \cite{bommeli} 
yields $\lambda_{tot}\approx 0.92$ rather close to the value 
of 1.05 derived from specific
heat data mentioned above. Similarly using $\Omega_{p}=6.01$~eV 
and $\delta=0.7$ \cite{bommeli} one arrives at $\lambda=0.996$
even still closer to the specific heat value.
Next, let us consider
 
(ii)
a typical A15-superconductor, namely,
V$_3$Si ($T_c=17$~K). Here $\lambda_L(0)\approx \lambda_L(T=3.8\mbox{K})=$ 
108~nm derived from $\mu$SR measurements by Sonier {\it et al.} 
\cite{sonier}. The experimental plasma frequency of
$\Omega_{pl}=$2.8 $\pm$ 0.2 ~eV \cite{mattheiss,yao}
has been obtained from room-temperature infrared data in between 0.16 
and 0.62~eV using a pure Drude-model relation according to it the slope 
$1/(1-\varepsilon_1)$ vs $\omega^2$ is given by $(\Omega_p/\hbar)^{-2}$.
But this means that no $\varepsilon_{\infty}$ has been taken into account.
Hence an $\Omega_p$ obtained this way actually represents 
a {\it screened} plasma frequency and the real unscreened plasma frequency
must be larger. For this reason one obtains too small disorder parameters
$\delta$ if one one adopts for total coupling {\it el-boson} constant 
$\lambda_{tot}=0.9$
obtained from an Eliashberg-theory analysis of its thermodynamic properties 
\cite{mitrovic} based on tunneling data for 
the spectral density $\alpha^2F(\omega)$ \cite{kihlstrom}. This 
procedure is the most 
accurate way to determine the relevant coupling strength for a 
standard superconductor. 
Then using the above mentioned penetration depth one would have
$(1+\lambda)(1+\delta)=$2.026, 2.35 to 2.7, 
i.e.\ $\delta=0.026, 0.24$ to 0.42 for the lower bound,
the mean value and the upper bound, respectively. Thus it is clear why
the lower bound is  unrealistic. We may refine this way 
$\Omega_p=2.9 \pm 0.1$. Then one is left with a ratio of 1.15  between
$\Omega_{p,{\rm LDA}}= $3.35~eV and  the refined experimental value. This
ratio is 
a bit smaller than $\sim 2$ found
for the pnictide superconductors considered above.
Another phonon mediated superconductor with a sizable {\it el-ph} 
coupling is

(iii) MgCNi$_3$ for which
$\lambda\approx 1.8$ has been found and $\Omega_{p, LDA}=3.17$ eV has 
been calculated 
 applying the mentioned above FPLO-code \cite{waelte}. 
According to table 2, and according to our general experience
for empirical plasma frequencies, one has $\Omega_p \approx (2/3) 
\Omega_{p,LDA}$. Thus we adopt $\Omega_p=2.1$~eV. Using the measured
penetration depth of $\lambda_T(0)=231.5$~nm \cite{macdougall}
one arrives at $DZ=6.07$ substituting our empirical strong coupling
value $\lambda=1.8$ 
we are left with a relative small disorder
parameter $\delta=1.17\sim 1$
 in accord with our previous assignment of a quasi-clean limit 
regime in this compound \cite{waelte}.
 
(iv) The Ba$_{1-x}$K$_x$BiO$_3$ ($T_c\approx$ 30 at optimal doping) exhibits
a penetration depth of $\lambda_L(0)=198.5$ nm \cite{zhao} and 
$\lambda=1.4$ has been 
estimated by Zhao \cite{zhao}. 
With an unscreened plasma frequency of 
$\Omega_{pl}=$2.9 to 3~eV \cite{nagata,bozovic} one arrives at relatively 
large values
$\delta \approx$ 2.55 to 3.25, where in the last case 
$\lambda\approx \lambda_{tr} \approx $1
suggest by Nagata {\it et al.} from the high-$T$ resistivity data using
equation (2) has been suggested. In other words Ba$_{1-x}$K$_x$BiO$_3$
is a dirty strongly coupled superconductor.

Turning 
(v) to the heavy fermion superconductor 
PrOs$_4$Sn$_{12}$ ($T_c=1.85$~K) and to its non-4$f$ counter part 
LaOs$_4$Sn$_{12}$ ($T_c=0.74$~K). Their penetration depths amount 
$\lambda_L(0)=$ 353.4~nm and 470~nm, respectively. To the best of our 
knowledge
there are no optical data available. But theoretical calculation
predict very strongly coupled superconductivity with $\lambda \approx $ 
3.3 in the 
former case and 
weakly coupled superconductivity with $\lambda \approx 0.3 $ in the last case
\cite{eremin}. 
Here the strong coupling contribution is expected due to nonphononic 
quadropolar crystal field excitations.
If this picture is correct, one might predict lower
bounds for the unscreened plasma frequency: $\Omega_{p}>$ 1.16 eV in the 
former case and 0.479~eV in the latter case.
Finally, turning to cuprates, we consider first

(vi) the hole (under)doped HTSC YBa$_2$Cu$_3$O$_{6.67}$ 
with $\lambda_L(0)=178.4$~nm (see figure 6 of reference \cite{sonier07})
and $\Omega_{p}\approx$ 2 eV one arrives at 
$(1+\lambda)(1+\delta)=$3.27, i.e.\ 
strong coupling 
$\lambda_{tot}\sim 2$ in the quasi-clean limit regime which is in nice accord
with the ARPES data of the underdoped YBa$_2$Cu$_3$O$_{6.6}$ system
where $\lambda_{tot}=1.59$ has been derived from a kink in the 
dispersion-laws
of the quasiparticles \cite{borisenko2} and it was described by the authors
mainly to interaction with the magnetic resonance mode.
We close this subsection considering 

(vii) the electron-doped Nd$_{1.85}$Ce$_{0.15}$CuO$_{4-\delta}$
where the screened plasma frequency derived from the peak of the loss 
function amounts about 1.06~eV  and the plasma edge (minimum)
in the reflectivity data occurs near 1.3~eV \cite{alexander}. 
The volume of the unit cell ammounts $v=188.7$~\AA $3$ \cite{tarascon}.
Using the ionic polarizabilities like those mentioned in subsection 4.2
we arrive at an unscreened value of $\Omega_p$ in between 2.2 and 
2.7~eV.
Using equation (\ref{clausius-mossotti}) we estimate 
$\varepsilon_\infty\approx 4.51$  and this way 
2.25~eV~$\leq \Omega_p \leq$ 2.76~eV. Anlage {\it et al.} \cite{anlage}
measured $\lambda_{L}(0)= 105 \pm 20$\AA . Whereas
Luke {\it et al.} found in $\mu$SR measurements $\lambda_L(0)=230$~nm.
Then we finally estimate
$\lambda_{tot}(0) \leq 1$ for this cuprate with a $T_c\approx 20$~K, only.
Our estimated coupling constant $\lambda_{tot}$ is in accord with a recent 
theoretical estimate
of 0.7 by Cappelutti {\it et al.} \cite{cappelutti} based on the 
Holstein-$t$-$J$ model and of 0.8 by Park {\it et al.} \cite{park} based
on an analysis of their ARPES data.
Hence, electron-doped cuprates exhibit a similar behaviour as 
LiFeAs and NaFeAs with respect to the relation of
$T_c$ and the unscreened plasma frequency (see figure 9) and with respect
to the {\it el-boson} coupling constant with 
La-1111 iron based superconductors.

\subsection{Multiband effects}
\begin{figure}[t]
\hspace{0cm}
\begin{minipage}{8.6cm}
\includegraphics[width=8.5cm,angle=270]{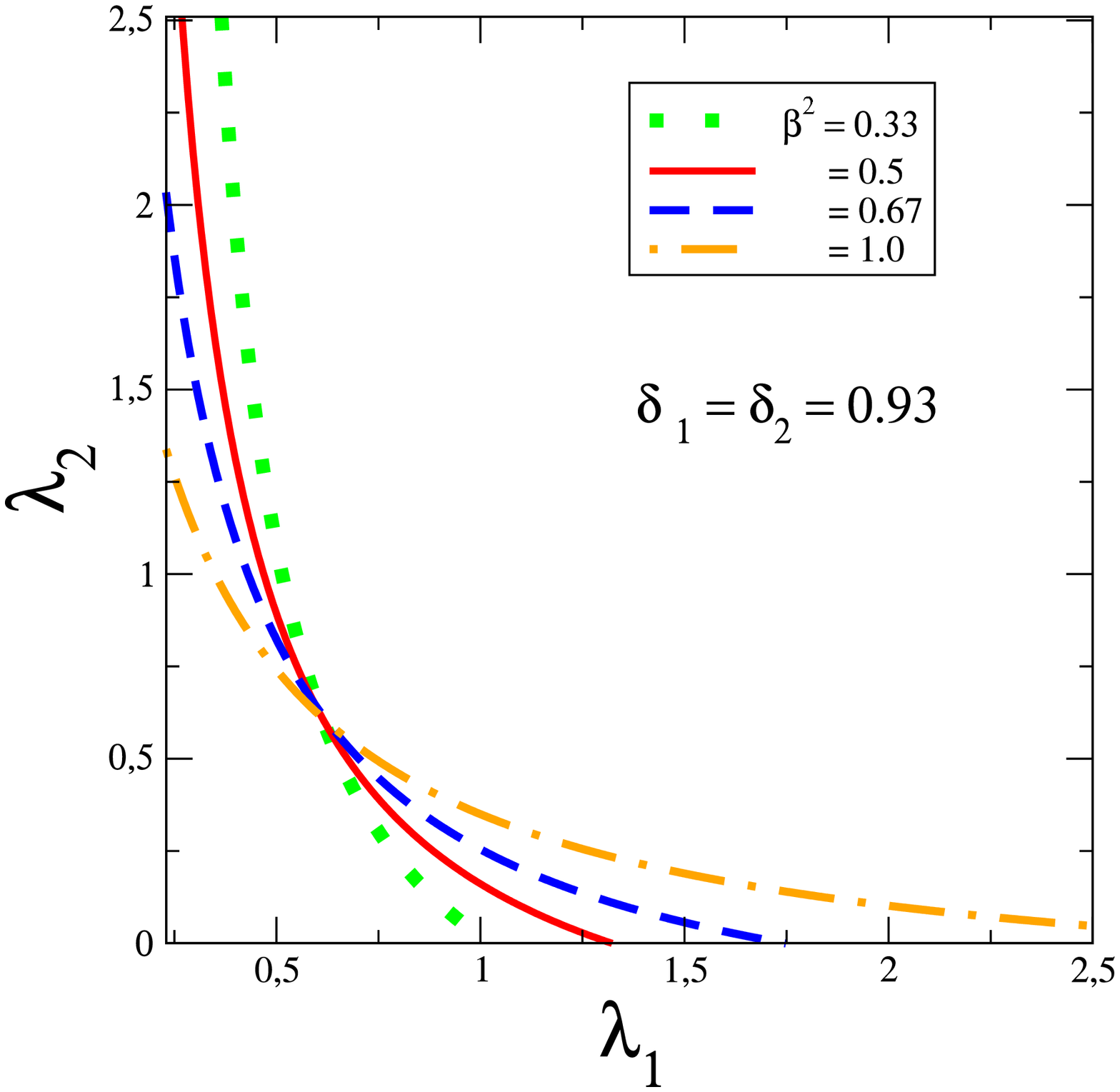}
\end{minipage}
\begin{minipage}{8.1cm}
\includegraphics[width=7.0cm]{fig10b.eps}
\end{minipage}
\caption{Constraint for the {\it el-boson} coupling constants 
$\lambda_1$ and $\lambda_2$ for
various individual
Drude plasma frequencies
measured by the parameter $\beta=\Omega_2/\Omega_1$ using the experimental 
value of the penetration depth $\lambda_L(0)=$254 nm \cite{luetkens} and the empirical total
unscreened plasma frequency $\Omega_p=$1.37 eV \cite{drechsler}
for LaO$_{0.9}$F$_{0.1}$FeAs (left) the same for enhanced disorder 
in the ``slow'' band 2 (right).
}
\label{twoband-lambda}
\end{figure}

Let us consider the simplest case of two effective bands, where for
instance one band could stand for the two electron bands and the second
band for the two hole bands predicted by the  DFT-LDA
calculations mentioned above. Rewriting the total unscreened plasma 
frequency of 
equation
(\ref{multiplasmon}) as
\begin{equation}
\Omega^2_{p,tot}=\Omega^2_1+\Omega^2_2\equiv\Omega^2_1\left(1+\beta^2\right),
\end{equation}
where $\beta =\Omega_2/\Omega_1$ and the index `p' has been ommitted, one 
obtains
the following straightforward generalization
of equation  (\ref{singlepen}) (compare also the two-band expressions
for the penetration depth as given in references \cite{golubov,nicol}):
\begin{equation}
\Omega_{1}\lambda_L(0)=\sqrt{\frac{Z_1D_1N_1}{1+\beta^2\frac{Z_1D_1N_1}{Z_2D_2N_2}}},
\label{multipen}
\end{equation}
where $Z_\alpha$ denotes
the mass enhancement factor of the band 
$\alpha=1,2$. $Z_{\alpha}= 1+\lambda_{\alpha}$ 
with the total coupling constant
(i.e.\ summed up over all
interacting {\it el-boson} channels)  
including also the 
interband scattering, $N_{\alpha} = n_{\alpha}/ n_{\alpha , s}$ is the 
inverse relative 
condensate occupation number of the
band $\alpha$ and $D_{\alpha}=1+\delta_{\alpha}$ is the 
corresponding disorder parameter
generalizing equation (\ref{delta}).
For coinciding coupling constants
and disorder parameters, i.e.\ $\lambda_1=\lambda_2$ and 
$\delta_1=\delta_2$,  the one-band picture with its weak or 
moderate coupling strength
discussed above is reproduced (see figures 8 and \ref{twoband-lambda}). Upon 
 weakening (strengthening) the coupling in one band say, in band 1,
stronger (weaker) couplings are introduced in band 2.  
Naturally, the  band with the larger coupling  exhibits the 
larger gap, if the corresponding 
couplings 
support the actual pairing state
which, however, has not been specified in our approach. 
From that constraint there is, in principle, taking into account the 
present poor knowledge of parameters no practical bound 
for $\lambda_2$, i.e.\ for the band with the 
slower 
quasi-particles, whereas the coupling constant 
$\lambda_1$ for the band with the 
faster charge carriers
is restricted from below and from above as 
well. At the upper bound $\lambda_2=0$, whereas it  formally diverges at 
the lower bound, provided 
$\lambda_1^{min}>0.$
In the special case when $\beta^2 \ll 1$, i.e.\ of a light and a heavy
band, the fast electrons may mask the heavy electrons. For the iron 
pnictides under
consideration
the hole bands do exhibit somewhat smaller Fermi velocities 
(and also plasma energies) according to band structure calculations. 
Therefore one might ascribe our effective band 2 to these hole pockets. 
In general, we regard equation (\ref{multipen}) as potentially helpful 
to identify
the location of the larger and the smaller
 gap values in addition to other methods.
A strongly coupled
superconducting heavy subgroup
which would be responsible for very high upper critical fields might
be this way
hidden by weakly coupled fast electrons.

\section{\bf Conclusion}
We performed a combined theoretical and experimental study of the screened 
and unscreened in-plane plasma frequencies of iron based pnictide 
superconductors. The obtained moderate renormalization of the corresponding 
optical masses is similar the one found frequently for other 
transition metals and is probably 
a normal 
Fermi liquid effect which gives strong support for the 
relevance of the
electronic structure
as calculated by various density functional based methods. 
Evidence for the presence of
weak correlation effects can be deduced also from the obtained large 
background dielectric constants, exceeding 11, which is ascribed to the 
large polarizability $\alpha_{As}$.
A comparison with zero temperature penetration 
depth 
data from muon spin rotation data provides a new constraint for the total
coupling strength of the {\it el-boson} interaction, the strength of impurity
scattering rates and the density of quasi-particles involved in the 
superconducting condensate. This information yields also reasonable 
estimates
for other more or less exotic superconductors. 
\ack
We thank Eschrig H, Mazin II, Koitzsch A, Borisenko SG, Shulga SV, 
Singh D,
Dolgov O, Kuli\'{c} M, Eremin I, Craco L, Luetkens H,
Plakida NM, Pickett WE, Timusk T, Maple B, Wang NL, van den Brink J, 
and Gvozdikov V
for discussions concerning various points of the present work 
and the DFG (DR 269/3-1 (S.-L.D), (M.K), the SFB 463 (S.-L.D, G.F., and H.R.) 
and its
Emmy-Noether Programme (H.R.)) and the ASCR project AVOZ10100520 
(J.M.) for financial
support.


\begin{thebibliography}{10}
\bibitem{remark}The experimentally observed unexpected 
negative value of $\alpha_{Fe^{2+}}$, if confirmed,  
is obviously a finite $\omega$-effect. The elucidation of the  
corresponding $\omega$-dependence of $\alpha_{Fe^{2+}}$ might be
also of interest for the pairing mechanism 
in iron based pnictides and related compounds
in addition to the large value
of $\alpha_{As}$ confirmed here. However, in the set recommended in
reference \cite{shannon} $\alpha_{Fe^{2+}}$ is positive.
\bibitem{remarkshannon}We note that 
an absolutely unreasonable large value of $\varepsilon_{\infty}= 14.82$ 
would be obtained,
if one would use equation (4) in reference \cite{shannon}. In the present
compound  for
the anionic polarizability of O$^{2-}$ only $v_c/10$ is 
available for one $O^{-2}$-ion. This 
would cause more than  a doubling of the oxygen polarizability:  
$\tilde{\alpha}_{{\rm O}^{2-}}=4.18$\AA$^3$ giving rise to the mentioned 
unrealistic value of $\varepsilon_{\infty}$ exceeding a realistic 
one by a factor of four.
\end{thebibliography}
\end{document}